\newcommand{\std}[1]{
  \usepackage{amssymb}
  \usepackage{amsmath}
  \usepackage[final]{graphicx}
  \usepackage{a4,a4wide}
\usepackage{maple2e}
  \renewcommand{\baselinestretch}{#1}
  \parindent 0pt
  \parskip 2ex
  \pagestyle{headings}
  \begin{document}
}
\newcommand{\article}[1]{
  \documentclass[12pt,fleqn]{article}\std{#1}
}
\newcommand{\artikel}[1]{
  \documentclass[12pt,twoside,fleqn]{article}\usepackage{german}\std{#1}
}
\newcommand{\revtex}{
  \documentstyle[preprint,aps,eqsecnum,amssymb,fleqn]{revtex}
  \begin{document}
}
\newcommand{\book}[1]{
  \documentclass[12pt,twoside,fleqn]{book}
  \addtolength{\oddsidemargin}{2mm}
  \addtolength{\evensidemargin}{-4mm}
  \std{#1}
}
\newcommand{\buch}[1]{
  \documentclass[10pt,twoside,fleqn]{book}
  \usepackage{german}
  \usepackage{makeidx}
  \makeindex
  \std{#1}
}
\newcommand{\foils}[1]{
  \documentclass[fleqn]{article}
  \usepackage{amssymb}
  \usepackage{amsmath}
  \usepackage[final]{graphicx}
  \renewcommand{\baselinestretch}{#1}
  \setlength{\voffset}{-3cm}
  \setlength{\hoffset}{-4cm}
  \setlength{\textheight}{27cm}
  \setlength{\textwidth}{19cm}
  \parindent 0ex
  \parskip 0ex 
  \pagestyle{plain}
  \begin{document}
  \huge
}
\newcommand{\landfoils}[1]{
  \documentclass[fleqn]{article}
  \usepackage{amssymb}
  \usepackage{amsmath}
  \usepackage[final]{graphicx}
  \renewcommand{\baselinestretch}{#1}
  \setlength{\hoffset}{-5cm}
  \setlength{\voffset}{-1.5cm}
  \setlength{\textwidth}{27cm}
  \setlength{\textheight}{19cm}
  \parindent 0ex
  \parskip 0ex 
  \pagestyle{plain}
  \begin{document}
  \huge
}

\newcommand{\horline}{
\vspace{-1ex}
\begin{list}{}{\leftmargin0ex \rightmargin0ex \topsep0ex}\item[]\hrulefill\end{list}
\vspace{1ex}}

\newcommand{\addressCologne}{
Institute for Theoretical Physics\\
University of Cologne\\
50923 K\"oln---Germany\\
{\tt mt@thp.uni-koeln.de}\\
{\tt www.thp.uni-koeln.de/\~{}mt/}
}

\newcommand{\homepage}{{\tt www.neuroinformatik.ruhr-uni-bochum.de/PEOPLE/mt/}}
\newcommand{\email}{{mt@neuroinformatik.ruhr-uni-bochum.de}}

\newcommand{\address}{
Institut f\"ur Neuroinformatik\\
Ruhr-Universit\"at Bochum, ND 04\\
44780 Bochum---Germany\\
\email
}

\renewcommand{\author}{{\bf Marc Toussaint} -- November, 1999\medskip\\ \addressCologne}

\newcommand{\mytitle}[1]{
\thispagestyle{empty}
\horline
\vspace{0ex}
\begin{list}{}{\leftmargin2ex \rightmargin5ex \topsep3ex }\item[]
{\huge\bf #1}
\end{list}
\begin{list}{}{\leftmargin7ex \rightmargin5ex \topsep0ex }\item[]
\author
\end{list}
\horline
}

\newcommand{\headline}[1]{\footnotetext{\sc Marc Toussaint, \today
    \hspace{\fill} file: #1}}

\newcommand{\sepline}{
\begin{center} \begin{picture}(200,0)
  \line(1,0){200}
\end{picture}\end{center}
}

\newcommand{\intro}[1]{\textbf{#1}\index{#1}}

\newtheorem{defi}{Definition}
\newenvironment{definition}{
\begin{quote}
\begin{defi}
}{
\end{defi}
\end{quote}
}

\newenvironment{block}{
\begin{quote} \begin{picture}(0,0)
        \put(-5,0){\line(1,0){20}}
        \put(-5,0){\line(0,-1){20}}
\end{picture}

}{

\begin{picture}(0,0)
        \put(-5,5){\line(1,0){20}}
        \put(-5,5){\line(0,1){20}}
\end{picture} \end{quote}
}

\newenvironment{summary}{
\begin{center}\begin{tabular}{|l|}
\hline
}{\\
\hline
\end{tabular}\end{center}
}

\newcommand{\inputReduce}[1]{
  
  {\sc\hspace{\fill} REDUCE file: #1}
}
\newcommand{\inputReduceInput}[1]{
  
  {\sc\hspace{\fill} REDUCE input - file: #1}
  \input{#1.tex}
}
\newcommand{\inputReduceOutput}[1]{
  
  {\sc\hspace{\fill} REDUCE output - file: #1}
}

\newcommand{\macros}{
  \newcommand{\0}{{\hat 0}}
  \newcommand{\1}{{\hat 1}}
  \newcommand{\2}{{\hat 2}}
  \newcommand{\3}{{\hat 3}}
  \newcommand{\5}{{\hat 5}}
  \newcommand{\QQ}{{\cal Q}}

  \renewcommand{\a}{\alpha}
  \renewcommand{\b}{\beta}
  \renewcommand{\c}{\gamma}
  \renewcommand{\d}{\delta}
    \newcommand{\D}{\Delta}
    \newcommand{\e}{\epsilon}
    \newcommand{\g}{\gamma}
    \newcommand{\G}{\Gamma}
  \renewcommand{\l}{\lambda}
  \renewcommand{\L}{\Lambda}
    \newcommand{\m}{\mu}
    \newcommand{\n}{\nu}
    \newcommand{\N}{\nabla}
  \renewcommand{\k}{\kappa}
  \renewcommand{\o}{\omega}
  \renewcommand{\O}{\Omega}
    \newcommand{\p}{\varphi}
  \renewcommand{\r}{\varrho}
    \newcommand{\s}{\sigma}
    \newcommand{\Si}{\Sigma}
  \renewcommand{\t}{\theta}
    \newcommand{\T}{\Theta}
  \renewcommand{\v}{\vartheta}
    \newcommand{\Y}{\Upsilon}

  \newcommand{\C}{{\bf C}}
  \newcommand{\R}{{\bf R}}
  \newcommand{\Z}{{\bf Z}}

  \renewcommand{\AA}{{\cal A}}
  \newcommand{\GG}{{\cal G}}
  \newcommand{\TT}{{\cal T}}
  \newcommand{\EE}{{\cal E}}
  \newcommand{\HH}{{\cal H}}
  \newcommand{\II}{{\cal I}}
  \newcommand{\KK}{{\cal K}}
  \newcommand{\MM}{{\cal M}}
  \newcommand{\CC}{{\cal C}}
  \newcommand{\PP}{{\cal P}}
  \newcommand{\RR}{{\cal R}}
  \newcommand{\YY}{{\cal Y}}
  \newcommand{\SOSO}{{\cal SO}}
  \newcommand{\GLGL}{{\cal GL}}

  \newcommand{\NNN}{\mathbb{N}}
  \newcommand{\ZZZ}{\mathbb{Z}}
  \newcommand{\RRR}{\mathrm{I\!R}}
  \newcommand{\CCC}{\mathbb{C}}
  \newcommand{\one}{{\bf 1}}

  \newcommand{\<}{\langle}
  \renewcommand{\>}{\rangle}
  \newcommand{\tr}{{\rm tr}}
  \newcommand{\lag}{\mathcal{L}}
  \newcommand{\inn}{\rfloor}
  \newcommand{\lie}{\pounds}
  \newcommand{\speer}{\parbox{0.4ex}{\raisebox{0.8ex}{$\nearrow$}}}
  \renewcommand{\dag}{ {}^\dagger }
  \newcommand{\h}{{}^\star}
  \newcommand{\w}{\wedge}
  \newcommand{\ow}{\stackrel{\circ}\wedge}
  \newcommand{\feed}{\nonumber \\}
  \newcommand{\comma}{\; , \quad}
  \newcommand{\period}{\; . \quad}
  \newcommand{\del}{\partial}
  \newcommand{\point}{$\bullet~~$}
  \newcommand{\doubletilde}{
  ~ \raisebox{0.3ex}{$\widetilde {}$} \raisebox{0.6ex}{$\widetilde {}$} \!\!
  }
  \newcommand{\topcirc}{\parbox{0ex}{~\raisebox{2.5ex}{${}^\circ$}}}
  \newcommand{\sym}{\topcirc}

  \newcommand{\half}{\frac{1}{2}}
  \newcommand{\third}{\frac{1}{3}}
  \newcommand{\fourth}{\frac{1}{4}}

}

\newcommand{\tmp}{\fbox{?}}
\newcommand{\Label}[1]{\label{#1}\fbox{\tiny #1}}

\macros
\newcommand{\path}{./}
\newcommand{\basepath}{./}
\newcommand{\setpath}[1]{
  \renewcommand{\path}{#1}
  \renewcommand{\basepath}{#1}}
\newcommand{\pathinput}[2]{
  \renewcommand{\path}{\basepath #1}
  \input{\path #2} \renewcommand{\path}{\basepath}}

\newcounter{parac}
\newcommand{\para}{\refstepcounter{parac}[\emph{\arabic{parac}}]~~}
\newcommand{\Pref}[1]{[\emph{\ref{#1}}\,]}

\book{1}

 \def\emptyline{\vspace{12pt}}
\DefineParaStyle{Maple Output}
\DefineCharStyle{2D Math}
\DefineCharStyle{2D Output}

\newcommand{\zz}[1]{\vspace{-3ex}\begin{flushleft} #1 \end{flushleft}\vspace{-4ex}}
\newcommand{\spc}{\vspace{1ex}}
\newcommand{\verbon}{\vspace{-2.5ex}\footnotesize\begin{quote}}
\newcommand{\verboff}{\end{quote}\normalsize\vspace{-2.5ex}}
\newcommand{\tton}{\vspace{-2.5ex}\footnotesize\begin{quote}\tt}
\newcommand{\ttoff}{\end{quote}\normalsize\vspace{-2.5ex}}
\newcommand{\mapon}{\vspace{-2.5ex}\footnotesize\[}
\newcommand{\mapoff}{\]\normalsize\vspace{-2.5ex}}
\newcommand{\smallinput}[1]{\footnotesize\input{#1}\normalsize}

\mytitle{Lectures on Reduce and Maple\\ at UAM-I, Mexico
\footnote{see also
{\tt www.thp.uni-koeln.de/\~{}mt/work/1999mexico/}}}

\parskip 0ex
\tableofcontents
\parskip 2ex

\chapter*{Computer Algebra (CA)}
\addcontentsline{toc}{chapter}{Computer Algebra}

In order to get an idea what facilities CA provides it might be
sensible to ask what we expect from CA. I suggest that

\begin{quote}
  CA is supposed to {\bf solve or simplify any algebraic problem} that
  we can state explicitly.
\end{quote}

So, CA has to provide methods to simplify algebraic expression and to
solve equation systems. This, of course, is not enough because it also
has enable us to input our problem and has to generate some output of
the result (also graphical, plots, LaTeX, etc.). I want to
differentiate the needs for the input a little bit more: On the one
hand, CA has to offer some language or formalism that includes many
operators to form expressions ($+,~*,~\sin,~\int,~\sum$, etc.) and on
the other it should provide methods to make definitions, declarations,
substitutions, and assumptions. Thus, we split the facilities we expect
from CA into four groups:

\begin{itemize}
\item[A~~] operators to form expressions
\item[B~~] methods to make definitions, declarations, substitution,
  and assumptions
\item[C~~] methods to simplify and solve
\item[D~~] methods to generate output
\end{itemize}

The reason why I suggest this splitting of the commands is the
following: It seems that the discussion of group A dominates most
handbooks on CA only because it is by far the largest group of
commands and operators. I find though, that the commands in group B
and C are more elementary and important to understand whereas for the
commands in group A one can well refer to the online user's manual.
The commands in group D are also important for efficient work with CA.
This is why these lectures try to emphasize the commands in groups
B, C, and D a little bit more than usual.

\chapter{Reduce I}

\section{The simplification principle}

As stated, I think the simplification and solving methods of a CA
system to be most elementary and important. Hence we start to discuss
these in the case of Reduce.

Reduce is an input-output machine. One could say that {\bf all that
  Reduce does is to reformulate your input expression obeying certain
  rules!} One of these rules is, e.g., to execute all operators and
commands in the expression. For example, Reduce \emph{reformulates}
$1+1$ by executing the $+$ operator and answering $2$. Fortunately,
the rules for reformulation are such that - usually - {\bf this
  reformulation means a simplification.} These rules can be influenced
by the user: Either by switching on and off the \emph{rule switches}
or by introducing new rules.  Most important, Reduce offers the rule
switches collected in table \ref{Tswitches}.

\begin{table}[t]\center
\begin{tabular}{cc|l|c}
&switch & description & e.g.\ \\
\hline
\tt *&\tt allfac & factorize simple factors & $2x+2 \to 2(x+1)$ \\
&\tt div & divide by the denominator & $(x^2 + 2)/x \to x + 2/x$ \\
\tt *&\tt exp & expand all expressions & $(x+1)(x-1) \to x^2-1$ \\
\tt *&\tt mcd & make (common) denominators & $x^{-1} \to 1/x$ \\
\tt *&\tt lcm & cancel least common multiples & \\
&\tt gcd & cancel greatest common divisor & \\
&\tt rat & display as polynomial in {\tt factor} & $\frac{x+1}{x} \to 1 + x^{-1}$ \\
\tt *&\tt ratpri & display rationals as fraction & $1/x \to \frac{1}{x}$ \\
\tt *&\tt pri & dominates {\tt allfac, div, rat, revpri} & \\
&\tt revpri & display polynomials in opposite order & $x^2+x+1 \to 1+x+x^2$ \\
&\tt rounded & calulate with floats & $1/3 \to 0.333333333333$ \\
&\tt complex & simplify complex expressions & $1/i \to -i$ \\
&\tt nero & don't display zero results & $0 \to ~~$ \\
\tt *&\tt nat & display in Reduce input format & $\frac{x^2}{3} \to x${\tt **}2/3\\
&\tt msg & suppress messages & \\
&\tt fort & display in Fortran format & \\
&\tt tex & display in TeX format & \\
\end{tabular}
\caption{Switches for Reduce's reformulation rules. Those marked with {\tt *} are turned on at default.}
\label{Tswitches}
\end{table}

As one can see, by switching on the {\tt exp}-rule, Reduce
reformulates and thus \emph{simplifies} the expression $(x+1)(x-1)$ as
$x^2-1$. This principle is quite different to other CA systems. The
combination of such rules can be very powerful. The rule switches
turned on at default are: {\tt allfac, exp, mcd, lcm, ratpri, pri,
  nat}.  This means that, at default, for any expression
(subexpression, component, etc. ), Reduce expands the multiplication
of two larger terms, interprets all divisions as rationals, tries to
divide common factors in numerators and denominators, and finally
tries to factorize simple factors. This is already quite a powerful
simplification scheme.

\section{The interface}

Starting Reduce on Unix it displays about

\verbon\begin{verbatim}
Loading image file:
  /vol/sc/lib/reduce3.6_patch980830/reduce.img REDUCE 3.6, 15-Jul-95,
  patched to 30 Aug 98 ...

1: 
\end{verbatim}
\verboff

The first three lines are an artifact of Reduce being implemented in
LISP: The program code (\emph{image}) of Reduce is passed to the LISP
interpretor. The prompt {\tt 1:~} expects a command as input. {\bf
  Reduce ignores all upper case letters!} A command consists out of a
statement and a terminator. The terminator decides whether the
Reduce's response is displayed ({\tt ;} terminator) or not ({\tt \$}
terminator). The following lines are command lines only, the response
of Reduce is (sometimes) indicated after the comments sign {\tt \%}.
All examples in this chapter are collected in the file {\tt red1} at
\cite{mypage}.

\verbon\begin{verbatim}
% file "red1"

off factor, exp, mcd, ratpri$
(x+1)**3/3 - x;                %-> 1/3*(x + 1)**3 - x
on mcd$          ws;           %-> ((x + 1)**3 - 3*x)/3
on ratpri$       ws;           %-> 
on exp$          ws;           %-> (x**3 + 3*x**2 + 1)/3
factor x$on rat$ ws;           %-> x**3/3 + x**2 + 1/3
remfac x$off exp$

z+a;                           %-> a + z
order z,a$ ws;                 %-> z + a
order nil$ ws;                 %-> a + z
\end{verbatim}
\verboff

\section{User definitions}

\subsection{Names and assignments}\label{Reducedefs}

\verbon\begin{verbatim}
off mcd$
my_name123 := (x**2-1)/(x+1);  %-> my_name123 := (x + 1)**(-1)*(x**2 - 1)
on mcd$ my_name123;            %-> x - 1
clear my_name123;
\end{verbatim}
\verboff

A string can be used as a \emph{name} (or identifier) for variables,
procedures, etc. The first character of a name has to be alphabetic,
whereas others can be numbers or '{\tt \_}', e.g. '{\tt a11}' and
'{\tt ric\_scalar0}' are proper names. A name should not coincide with
a reserved variable name ({\tt e i infinity nil pi t}) or the command
names in table \ref{TReducecoms}.

As long as a name is not assigned to some (symbolic) value it is
considered to be a \emph{clear name}. The assignment operator

\tton
{\it name}:={\it expr};
\ttoff

assigns the (reformulated) expression on the rhs to the name on the
lhs. To withdraw an assignment one uses the command

\tton
clear {\it name};
\ttoff

This command also deallocates the memory resource that is associated
with the name.

{\bf For understanding Reduce, it is important to distinguish between
  the expression assigned to a name and the value that Reduce displays
  when evaluating the name.} The last of which was reformulated by
Reduce with current rules and definitions:

\verbon\begin{verbatim}
a:=b$ a;                       %-> b
b:=1$ a;                       %-> 1
a:=a$b:=2$ a;                  %-> 1
clear a,b;
\end{verbatim}
\verboff

In the first line, {\tt a} is assigned to {\tt b} and is, of course,
also evaluated to be {\tt b}. In the second line, {\tt a} is still
assigned to {\tt b} but is evaluated to be {\tt 1} (because {\tt b} is
assigned to {\tt 1}). In the third line, {\tt a} is assigned to {\tt
  1} (because the rhs {\tt a} is evaluated to be 1) and an thus it is
also evaluated to be {\tt 1} even after {\tt b} is assigned to {\tt
  2}. We see that the reassignment ({\tt a:=a\$}) is not trivial at
all! Analogously, the following example shows the importance of the
reassignment if rules have changed:

\verbon\begin{verbatim}
off mcd$ a:=(x**2-1) / (x-1);     %-> a := (x**2 - 1)*(x - 1)**(-1)
on mcd$  a := a$  a;              %-> x + 1
off mcd$ a;                       %-> x + 1
\end{verbatim}
\verboff 

Here, in the end, {\tt a} is assigned to {\tt x+1}. Without the
reassignment {\tt a:=a\$}, {\tt a} would still be assigned to {\tt
  (x**2 - 1)*(x - 1)**(-1)} and also evaluated to be such in the last
line.

{\bf The reassignment is one of the most important tools to apply new
  definitions and to use the simplification methods of Reduce}.

\subsection{Aliases}

To define aliases for any expression, one can use the {\tt define}
command. The syntax is

\tton
define {\it alias}={\it expression};
\ttoff

Note that the rhs expression is evaluated before it is assigned to the
alias.  However: {\bf An alias can never ever be changed or unassigned
  during a Reduce session!} If an alias appears in an expression it is
replaced by its associated value before Reduce applies any other
rules.  Example:

\verbon\begin{verbatim}
define isprime = primep;
for x:=0:20 do if isprime(x) then write x," is a prime!";
                                       % throws 2 3 5 7 11 13 17 19

define is_assigned_to = :=;
x is_assigned_to 5;                    %-> x := 5
\end{verbatim}
\verboff

\subsection{Substitutions}

Actually, we already noticed how to substitute identities into
expressions: Define the identity with an assignment statement or
introduce a rule for this identity. Then reevaluate the expression.
Say, e.g., we want to substitute $x=0$ into $f(x)=cos(x)$, then we
could write one of the following three lines:

\verbon\begin{verbatim}
f := cos(x)$ x:=0$            f; clear x$        %-> 1
f := cos(x)$ let { x=>0 }$    f; clear x$        %-> 1
f := cos(x)$ f where { x=>0 };                   %-> 1
f := cos(x)$ sub(x=0,f);                         %-> 1
\end{verbatim}
\verboff

The last possibility is the most beautiful, because one does not have
to clear the assignment {\tt x:=0} or the rule {\tt x=>0} hereafter.
The {\tt where} command only introduces a local rule. After these
substitutions, {\tt f} is still assigned to {\tt cos(x)} but was only
\emph{locally} evaluated to be {\tt 1}. Reduce also offers the {\tt
  match} command to substitute expressions for polynomials instead of
names.

\subsection{Rules}

In the upper example we already introduced the {\tt let} command that
introduces new rules. In general, a rule has syntax

\tton
{\it expr} => {\it expr} [when {\it boolean}]
\ttoff

and the {\tt let} command accepts a list of rules as parameter. Rules
can be of explicit nature where any literal occurrence of the lhs
expression is replaced by the rhs expression. Or they can by of
parametric nature where the lhs expression includes formal names (like
parameters) that represent any expressions. These formal names are
marked with a twiddle such as {\tt \~{}x}:

\verbon\begin{verbatim}
sin(x)**2 + cos(x)**2;         %-> cos(x)**2 + sin(x)**2
trig_rules:={
 cos(~x)*cos(~y) => (1/2)*(cos(x+y)+cos(x-y)),
 sin(~x)**2 + cos(~x)**2 => 1 }$
let trig_rules$
sin(x+1)**2 + cos(x+1)**2;     %-> 1
cos(pi/3-1) * cos(1);          %-> (2*cos((pi - 6)/3) + 1)/4
showrules cos;                 % shows ALL rules associated with cos!
clearrules trig_rules;
\end{verbatim}
\verboff

As already mentioned, the {\tt where} command applies rules only one
the preceding expression. The command

\tton
showrules {\it name};
\ttoff

displays \emph{all} rules associated with the name. This gives an
insight in how Reduce defines functionals:

\verbon\begin{verbatim}
1: showrules log;

{log(1) => 0,

 log(e) => 1,

      ~x
 log(e  ) => x,

                    1
 df(log(~x),~x) => ---}
                    x
\end{verbatim}
\verboff

\subsection{Operators}

An operator is any name that accepts parameters. In Reduce, one can
declare a name to be an operator without at all specifying how many
parameters the operator expects and what the functionality of the
operator is! This leaves the user large freedom to handle and use
operators. E.g., one can introduce indexed coordinates $x^i$ by
declaring $x$ to be an operator with the index as only parameter and
without any further properties. In fact, most other, more specialized
declarations (like matrix, procedure, etc.) can be thought of special
kinds of operator declarations. And also functions (such as sin, log,
etc.) can be introduced as operators with certain properties. The
properties of an operator are declared by specifying the rules how
Reduce has to handle them. Example:

\verbon\begin{verbatim}
operator eps$                  % declares a new operator
antisymmetric eps$             % declares eps to be totally antisymmetric
let { eps(0,1,2,3) => 1  }$    % defines the properties of this operator
eps(3,1,2,0); eps(2,0,1,1);    %-> -1 and 0
showrules eps;                 %-> {eps(3,2,1,0) => 1}$
\end{verbatim}
\verboff

\verbon\begin{verbatim}
clear my_sin$ operator my_sin$
my_sin_rules:={
  my_sin(~x) => - my_sin(-x)        when numberp(x/pi) and x/pi<0,
  my_sin(~x) =>   my_sin(x - 2*pi)  when numberp(x/pi) and x/pi>2,
  my_sin(~x) => - my_sin(x - pi)    when numberp(x/pi) and x/pi>1,
  my_sin(~x) =>   my_sin(pi - x)    when numberp(x/pi) and x/pi>1/2,
  my_sin(0)    => 0,
  my_sin(pi/6) => 1/2,
  my_sin(pi/4) => 1/2*sqrt(2),
  my_sin(pi/3) => 1/2*sqrt(3),
  my_sin(pi/2) => 1
}$
let my_sin_rules$
my_sin(-124*pi/6);             %-> - sqrt(3)/2
clearrules my_sin_rules;
\end{verbatim}
\verboff

\section{Solving}

\verbon\begin{verbatim}
solve(x**3-3*x**2-61*x+63,x);  %-> {x=9,x=1,x=-7} which are the nulls of this polynomial
solve({x+y-9,-2*y**2+3*x},{x,y});   %-> { {y= - 9/2,x=27/2} , {y=3,x=6} }
\end{verbatim}
\verboff

Reduce tries to solve algebraic equation systems with

\tton
solve({\it expr\_list},{\it name\_list});
\ttoff

Without the switch {\tt multiplicities} Reduce display multiple
solutions only once.

Reduce has no build-in routines to solve differnential equations.
However, there exist packages for this problem. One should especially
mention the CATHODE 2 project (Computer Algebra Tools for Handling
Ordinary Differential Equations) which develops routines to hande
differential equation systems also for Reduce. The members of this
group are M.\ MacCallum (London), V.\ Fairen (Madrid), E.\ Tournier
(Grenoble), Th.\ Mulders (Zurich), M.\ van der Puf (Groningen),
F.\ Schwarz (Bonn), L.\ Brenig (Brusssel). You find their packages at
\cite{cathodepage}, especially {\bf Crack} and {\bf ODEsolve}.

\section{Commands and references}

Most of the commands Reduce offers are to formulate expressions. I
tried to collect most in table \ref{TReducecoms}. The point of this
table is to give you an idea what commands there exist and to briefly
describe the syntax. One should read through this table once. For
detailed explanations we refer to the user's manual which can be
accessed in the internet at \cite{reducepage}. I prepared more online
references at \cite{mypage}.

\begin{table}
\begin{tabular}{p{13ex}|p{74ex}}

control & \tt quit showtime pause cont;~~ ws input {\it prompt\_number};~~
\newline write {\it exprs};~~ in out shut load\_package "{\it
  filename}";~~ \newline rederr "{\it message}";~~ \% {\it comment} \spc\\ 

assignment & \tt {\it name} := {\it expr};~~ set({\it name},{\it
  expr});~~ define {\it name}:={\it expr};~~ \spc\\ 

unassignment & \tt clear {\it names};~~ \spc\\ 

substitutions & \tt let {\it rule\_list};~~ sub match ({\it eqn\_list},{\it
  expr});~~ {\it expr} where {\it rule\_list};~~\spc\\ 

rules & \tt {\it expr} => {\it expr};~~ let clearrules {\it rule\_list};~~
showrules {\it name};~~
\spc\\ 

elementary & \tt {\it expr} + - * / ** {\it expr};~~ \spc\\ 

logical & \tt {\it expr} = neq > >= <= < {\it expr};~~ ordp ({\it
  names});~~\newline {\it boolean} and or {\it boolean};~~ not {\it
  boolean};\newline numberp fixp evenp primep ({\it expr});~~ freeof
({\it expr},{\it name}); \spc\\ 

selection & \tt rhs lhs {\it eqn};~~ num den {\it expr};~~ coeff ({\it
  expr},{\it name});~~ \newline coeffn ({\it expr},{\it name},{\it
  degree});~~ \spc\\ 

functional & {\tt cos sin tan csc sec cot a\~{} \~{}h a\~{}h
  atan2 abs sqrt exp ln log logb log10 hypot factorial fix
  ceiling floor nextprime conj impart repart random round sign ({\it
    expr});~~ max min ({\it expr\_list});~~} \spc\\ 

calculus & \tt int df ({\it expr},{\it names});~~ depend nodepend ({\it name,names});\spc\\ 

matrix & \tt mat ({\it components});~~ tp trace det rank mateigen ({\it
  matrix});~~ \spc\\ 

lists & \tt \{{\it elements}\};~~ first second third rest reverse
length ({\it list});~~ \newline append ({\it list},{\it list});~~ {\it
  expr} .\ cons {\it list};~~ \spc\\ 

loops & \tt for {\it (see section \ref{ReduceLoops})};~~ while {\it
  boolean} do {\it statement};~~ \newline repeat {\it statement} until
{\it boolean};~~ \spc\\ 

conditions & \tt if {\it boolean} then {\it statement} [else {\it statement}];~~ \spc\\

groups & \tt begin scalar {\it names}\$ {\it statements} return {\it
  expr}\$ end;~~ \newline << {\it statements} >>;~~ \spc\\ 

ordering & \tt order korder factor {\it names};~~ \spc\\

solving & \tt solve ({\it expr\_list},{\it name\_list});~~\spc\\

declarations & \tt operator matrix {\it names};~~ array {\it
  name}({\it dimensions});~~ \newline procedure {\it name}({\it
  parameter\_names})\$ {\it statement};~~ \spc\\ 

properties & \tt noncom symmetric antisymmetric even odd linear {\it operators};~~\spc\\
\end{tabular}
\caption{
  Standard commands in Reduce. {\bf Explanation:} Each group of
  commands which are only separated by blanks have the same syntax.
  These groups are separated by '{\tt ;}' and the syntax in specified
  only once for all commands in one group (think of the blanks as
  'or'). The plural of italic words means that there can be many of
  them separated by commas. Terms in brackets '[ ]' are optional. The
  twidle {\tt \~{}} means one of {\tt sin, cos,} etc. {\it
    expr}=expression, {\it eqn}=equation (which has syntax {\tt {\it
      expr}={\it expr}}) }
\label{TReducecoms}
\end{table}

\verbon\begin{verbatim}
for l:=1:50 sum l;             %-> 1275
for l:=1:100 product l;        % gives the huge number 100!

int(sin(x)**2,x);                 %-> x/2 + (-cos(x)*sin(x))/2
depend (f,x)$ df(f,x); int(f,x);  %-> df(f,x)  int(f,x)
nodepend (f,x)$df(f,x);int(f,x);  %->    0       x*f

matrix m(3,2)$                 % declare matrix
m:=mat((1,2,3),(2,1,2),(3,2,1))$ % construct matrix
det m;                         %->8 determinante
1/m;                           % gives the inverse
matrix v(3,1)$                 % declare a vector
v:=mat((1),(2),(3))$
1/m * v;                       %->(1,0,0) matrix multiplication

procedure fac1(k)$
  for l:=1:k product l$        % defines a new factorial procedure:
fac1(16);                      %-> 20922789888000 which is 16!

operator fac2$
let { fac2(1) => 1 ,
      fac2(~n) => n*fac2(n-1) when numberp n and n > 1}$
fac2(16);                      %-> 20922789888000

write "next ""prime > 500"" is ",nextprime(500)$ %-> next "prime > 500" is 503

showtime$                      % milli seconds since last showtime
in "my_file"$                  % executes all lines in "my_file" as input
quit$                          % exits Reduce
\end{verbatim}
\verboff

\section{Exercises}

\begin{enumerate}
\item Get used to Reduce's interactive interface. Try control
  commands like {\tt showtime write quit input ws clear \%}.

\item Turn all switches (for the reformulation rules) off:

\tton
 off allfac,div,exp,mcd,lcd,ratpri,nat\$
\ttoff

Now type in term {\tt (x**2 - 1)/((x+1)*(x-1));} Which switches do you
have to turn on additionally to get the following reformulations by
Reduce:

a) {\tt (x**2 - 1)*(x + 1)**(-1)*(x - 1)**(-1)\$}

b) {\tt (x**2 - 1)**(-1)*x**2 - (x**2 - 1)**(-1)\$}

c) {\tt (x**2 - 1)**(-1)*(x**2 - 1)\$}

d) {\tt 1\$}

We see that {\tt mcd} is absolutely neccessary to calculate with
rationals. The {\tt exp} switch is important when it is not obvious
that terms cancel in large expressions. Similary, the {\tt gcd} switch
improves the perfromance in canceling a common divisor of numerator and
denominator (which {\tt mcd} does also in easier cases).

\item What is the value of
\begin{align}
a(a+2) + c(c-2) - 2ac
\end{align}
for $a-c=7$?

\item Get used to the user's manual at\\ 
  \verb+http://www.uni-koeln.de/REDUCE/3.6/doc/reduce/reduce.html+ and
  to the file execution with the {\tt in} command (See also section
  \ref{incom}). Find out the syntax of the commands {\tt limit
    factorial int df for}.  Thereby get answers for
\begin{align}
\lim_{x\to 0} \frac{sin(x)}{x} \comma 15! \comma \int\! f'(x)\, dx \comma \int\! \exp(f(x))\, f'(x)\, dx \comma 1-\sum_{i=1}^{10} \frac{1}{2^i} \;.\nonumber
\end{align}

Generate the list $\{2^0,2^1,..,2^{10}\}$ (use {\tt for} and {\tt collect}).

\item Does Reduce confirm
\begin{align*}
\left(\frac{\del z}{\del x}\right)^2 + \left(\frac{\del z}{\del y}\right)^2 = a e^{-z} \;?
\end{align*}
Teach Reduce the appropriate rule for such problems!

\item Generate a list of all prime numbers smaller than 1000 and being
  a divisor of 606353.

\item Write a procedure that counts the zeros at the end of the number
  $k!$ for any positive integer $k$.

\item Develop a procedure to determine the $n$th order Taylor
  expansion of some arbitrary function $f(x)$ at one arbitrary point
  $x=x_0$. Test the procedure by considering $e^x$ and $\sin x$.
  Confirm $e^{iz}=\cos z + i \sin z$ up to 5th order.

\item Prove by complete induction that
\begin{align*}
\sum_{k=2}^n\,\frac{1}{(k-1)\,k} = \frac{n-1}{n} \;.
\end{align*}

\item Write a procedure that return the characteristic polynomial and
  the eigen values of a square matrix. (Check with but don't use {\tt
    mateigen}.) Implement operators that return the scalar and vector
  product, $a \cdot b$ and $a \times b$, for two vectors $a$, $b$. Use
  arrays for these problems if you want indices to run from 0.

\end{enumerate}

\chapter{Reduce II}

\section{Running files with Reduce and packages}\label{incom}

For most purposes it is convenient to write all commands to be
executed by Reduce into a separate file edited with your favorite
editor. The {\tt in} command causes Reduces to run this file as if all
the lines where typed in during an interactive session. Use {\tt in
  "{\it filename}";} if you want Reduce to display each line of the
file and {\tt in "{\it filename}"\$} if not. The last line of such a
file should be {\tt end\$}. Other useful command in this context are
{\tt pause, cont, demo}.

On Unix systems one can also use the following command line to pipe
your Reduce file {\tt sample.rei} into reduce and collect all output
in the file {\tt sample.reo}\

\tton
reduce < sample.rei > sample.reo
\ttoff

Just like running own file you can read \emph{packages} that implement
new commands or whole calculi with the command {\tt load\_package}.
Table \ref{Reducepackages} briefly displays the packages available for
Reduce. In principle, one can think of packages as ordinary Reduce
code which is \emph{precompiled into a fast loading file}. (Such files
can be generated with the commands {\tt faslout "{\it filename}";
  faslend;}).  Later, we will describe the Excalc package implementing
the exterior calculus. You find a list of all packages and links to
online references in the table of contents of the Reduce user's manual
at \cite{reducepage}

Also, the ZIB (Berlin, Germany) offers a set of references (in PDF
format) for all packages at \cite{zibpackages}.

\begin{table}\footnotesize
\mbox{~}\hspace{-8ex}\begin{tabular}{p{15ex}|p{100ex}}
ALGINT & integration for functions involving roots (James H. Davenport) \spc\\
ARNUM & algebraic numbers (Eberhard Schrüfer) \spc\\
ASSIST & useful utilities for various applications (Hubert Caprasse) \spc\\
AVECTOR & vector algebra (David Harper) \spc\\
CALI & package for computational commutative algebra (Hans-Gert Graebe) \spc\\
CAMAL & calculations in celestial mechanics (John Fitch) \spc\\
CHANGEVAR & transformation of variables in differential equations (G. Üçoluk) \spc\\
COMPACT & condensing of expressions with polynomial side relations (Anthony C. Hearn) \spc\\
CRACK & package for solving overdetermined systems of PDEs or ODEs (Andreas Brand, Thomas Wolf)  \spc\\
CVIT & Dirac gamma matrices (V.Ilyin, A.Kryukov, A.Rodionov, A.Taranov) \spc\\
DESIR & differential equations and singularities (C. Dicrescenzo, F. Richard-Jung, E. Tournier) \spc\\
EXCALC & calculus for differential geometry (Eberhard Schrüfer) \spc\\
FIDE & code generation for finite difference schemes (Richard Liska) \spc\\
GENTRAN & code generation in FORTRAN, RATFOR, C (Barbara Gates) \spc\\
GNUPLOT & display of functions and surfaces (Herbert Melenk) \spc\\
GROEBNER & computation in multivariate polynomial ideals (Herbert Melenk, H.Michael Möller, Winfried Neun) \spc\\
HEPHYS & high energy physics (Anthony C. Hearn) \spc\\
IDEALS & Arithmetic for polynomial ideals (Herbert Melenk) \spc\\
LAPLACE & Laplace and inverse Laplace transform (C. Kazasov et al.) \spc\\
LIE & functions for the classification of real n-dimensional Lie algebras (Carsten, Franziska
     Schöbel) \spc\\
LIMITS & a package for finding limits (Stanley L. Kameny) \spc\\
LININEQ & linear inequalities and linear programming (Herbert Melenk) \spc\\
NUMERIC & solving numerical problems using rounded mode (Herbert Melenk) \spc\\
ODESOLVE & ordinary differential equations (Malcolm MacCallum et al.) \spc\\
ORTHOVEC & calculus for scalar and vector quantities (J.W. Eastwood) \spc\\
PHYSOP & additional support for non-commuting quantities (Mathias Warns) \spc\\
PM & general algebraic pattern matcher (Kevin McIsaac) \spc\\
REACTEQN & manipulation of chemical reaction systems (Herbert Melenk) \spc\\
RLFI, TRI  & TeX and LaTeX output (Richard Liska, Ladislav Drska, Werner Antweiler) \spc\\
ROOTS & roots of polynomials (Stanley L. Kameny) \spc\\
SCOPE & optimization of numerical programs (J. A. van Hulzen) \spc\\
SPDE & symmetry analysis for partial differential equations (Fritz Schwarz) \spc\\
SPECFN & package for special functions (Chris Cannam et al.) \spc\\
SPECFN2 & package for special special functions (Victor Adamchik, Winfried Neun) \spc\\
SYMMETRY & symmetry-adapted bases and block diagonal forms of symmetric matrices (Karin
     Gatermann) \spc\\
SUM & sum and product of series (Fuji Kako) \spc\\
TAYLOR & multivariate Taylor series (Rainer Schöpf) \spc\\
TPS & univariate Taylor series with indefinite order (Alan Barnes, Julian Padget) \spc\\
WU & Wu Algorithm for polynomial systems (Russell Bradford) \spc\\
\end{tabular}
\caption{Reduce packages}
\label{Reducepackages}
\end{table}

\section{Examples}

\subsection{Generating a Julia set}

\verbon\begin{verbatim}
% file "red3"

on comp$

procedure julia(c,s,file)$ begin
  on complex,rounded$
  precision(6)$
  l:={}$
  for x:=-3/2*s:3/2*s do << write x$ for y:=-3/2*s:3/2*s do <<
    z:=x/s+i*y/s$
    j:=0$ repeat z:=z**2+c until
      abs(repart(z))>2 or abs(impart(z))>2 or (j:=j+1)=50$
    if j=50 then l:={x,y}.l$
  >> >>$
  off complex,rounded$

  out file$
  for each point in l do
    write first(point)," ",second(point)$
  shut file$
end$

showtime$
julia(-0.11+0.67*i,100,"l2.red")$
showtime;

% time: 780730 ms

%in gnuplot:
%set data style dots
%set nokey; set noxtics; set noytics; set size square; set noborder
%plot "l2.red"
%set terminal postscript
%set output "julia1.ps"
%replot
\end{verbatim}
\verboff

\begin{figure}[t]\center
\includegraphics[scale=0.3,angle=0]{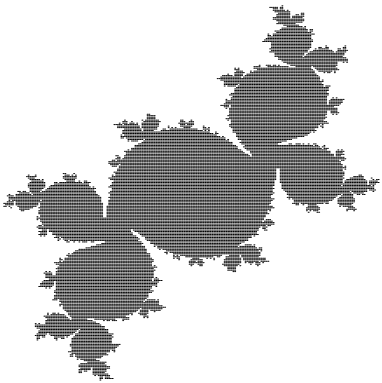}
\caption{The Julia set for $z \to z^2 + (-0.11\!+\!0.67i)$. (File \emph{julia1.ps}).}
\end{figure}

\subsection{Proving the Casimirs of the Poincar\'e group}

\verbon\begin{verbatim}
% file "red2"

% defining the epsilon and the Minkowsi metric:
operator eps$ antisymmetric eps$   
let eps(0,1,2,3)=>1$
array mink(3,3)$
mink(0,0):=1$ mink(1,1):=mink(2,2):=mink(3,3):=-1$ 

% declaring the generators of the Poincare group:
operator j,p$ noncom j,p$ antisymmetric j$ 

% the algebra of the Poincare group and general rules for the Lie bracket:
% note that the order of these rules is very important!
infix lie$
let {
  (  p(~a)  lie  p(~b)  ) => 0,
  (  p(~a)  lie j(~b,~c)) => mink(a,b)*p(c) - mink(a,c)*p(b),
  (j(~a,~b) lie  p(~c)  ) => - (p(c) lie j(a,b)),
  (j(~a,~b) lie j(~c,~d)) => - mink(a,c)*j(b,d) - mink(b,d)*j(a,c)
                             + mink(a,d)*j(b,c) + mink(b,c)*j(a,d),
  (- ~x lie ~y) => - (x lie y), ( ~x lie - ~y ) => - (x lie y),
  (~x lie ~y) => x * y - y * x
}$

% rules for applying commutator to order products of operators
l:=oplist:={p(0),p(1),p(2),p(3),j(0,1),j(0,2),j(0,3),j(2,3),j(3,1),j(1,2)}$
for each x in l do 
  for each y in (l:=rest l) do let y * x => x * y - (x lie y)$

% the momentum square (p2), the Pauli-Lubanski (pl), and its square (pl2):
operator p2,pl,pl2$
p2:=for a:=0:3 sum for b:=0:3 sum mink(a,b)*p(a)*p(b)$
for a:=0:3 do pl(a) := (-1/2)* 
  for b:=0:3 sum for c:=0:3 sum for d:=0:3 sum eps(a,b,c,d)*j(b,c)*p(d)$
pl2:=for a:=0:3 sum for b:=0:3 sum mink(a,b)*pl(a)*pl(b)$

% the commutators of p2 and pl2 with all generators:
for each x in oplist do write "[ p2 , ",x," ] = ",p2 lie x$
for each x in oplist do write "[ pl2 , ",x," ] = ",pl2 lie x;
\end{verbatim}
\verboff

\section{Advanced structures}

\subsection{Lists}

Lists are ordered sets. They are constructed via

\tton
\{ {\it elements} \};
\ttoff

The meaning of the commands to manipulate list should become clear
from the following example:

\verbon\begin{verbatim}
li:={a,b,c,d};
first li;                %-> a
second li;               %-> b
third li;                %-> c
part (li,4);             %-> d
rest li;                 %-> {b,c,d}
reverse li;              %-> {d,c,b,a}
length li;               %-> 4 
append (li,{e,f,g});     %-> {a,b,c,d,e,f,g} 
0 . li;                  %-> {0,a,b,c,d}
0 . 1 . 2 . 3 . li;      %-> {0,1,2,3,a,b,c,d}
\end{verbatim}
\verboff

\subsection{Loops and conditions}\label{ReduceLoops}

The syntax for the loop and conditional commands are

\tton
for {\it name} := {\it start}:{\it stop} [do|sum|product|collect|join] {\it statement}; \\
for {\it name} := {\it start} step {\it stepsize} until {\it stop} [do|sum|product|collect|join] {\it statement}; \\
for each {\it name} in {\it list} [do|sum|product|collect|join] {\it statement}; \\
while {\it boolean} do {\it statement}; \\
repeat {\it statement} until {\it boolean}; \\
if {\it boolean} then {\it statement};\\
if {\it boolean} then {\it statement} else {\it statement};\\
\ttoff

The {\tt do} action simple executes the statement in each iteration.
With the {\tt sum, product} actions, the hole {\tt for} statement
returns the sum or product of all statements in the iteration.
Analogously, the {\tt collect} action returns a list of all statements
and the {\tt join} action returns the union of all statements (these
have to be lists in this case!).

Examples:

\verbon\begin{verbatim}
for x:=1:5 do write x;                           % writes the numbers from 1 to 5
for each x in {2,4,8} sum x/2;                   %-> 7
for each x in {a,b,c} join {x,2*x};              %-> {a,2*a,b,2*b,c,2*c}
if x neq 0 and not x>0 then write "impossible!"; % returns nothing
\end{verbatim}
\verboff

As we can see, a {\it boolean} is an expression formed
by the logical operators {\tt = neq > >= <= < and or not}.

\subsection{Groups}

Instead of executing only \emph{one} statement in each loop or in some
conditional case, we can execute an arbitrary set of statements by
embracing them to a group. A simple way to form a group is

\tton
<< {\it 1st\_statement}; {\it 2nd\_statement}; ... ; {\it last\_statement} >>
\ttoff

This group is again a statement with the value of the {\it last
  statement} (if one did \emph{not} put a terminator after the last
statement). A more robust way to form groups is

\tton
begin [scalar {\it local\_names};] {\it statements} [return {\it value};] end;
\ttoff

Here, {\it local\_names} are such that have no effect outside of this
group. They are initialized with {\tt 0}. This group returns the
specified {\it value}. To produce an output within a group one has to
use the {\tt write} command.

\subsection{Procedures}

Procedures are operators with an explicitly defined set of parameters
and an explicitly defined functionality. The syntax for defining
procedures is:

\tton
procedure {\it name} ({\it parameter\_names}); {\it statement};
\ttoff

Here, {\it statement} defines the (return) value of the procedure and
is usually a group statement.

\subsection{Operators}

Operators can be declared prefix or infix by:

\tton
operator {\it names};\\
infix {\it names};
\ttoff

The declaration

\tton
precedence {\it name},{\it next\_lower\_precedence\_operator};
\ttoff

specifies the precedence of this operator: The declared operator
has just higher precedence than the operator specified.

Operators can have one of the following properties:\\
$~~~\cdot~$totally {\tt symmetric} or {\tt antisymmetric},\\
$~~~\cdot~${\tt even} or {\tt odd} with respect to parity of the parameter,\\
$~~~\cdot~${\tt linear}, or\\
$~~~\cdot~${\tt noncom}-muting under the multiplication {\tt *}\\
All of these properties can be declared by

\tton
{\it property} {\it operator\_names};
\ttoff

where {\it property} is one of {\tt symmetric, antisymmetric, even,
  odd, linear, noncom}.

\subsection{Arrays, matrices}

For handling multicomponent objects Reduce offers the following
declarations:

\tton
array({\it dimensions-1});\\
matrix({\it co\_dimension},{\it contra\_dimension});\\
\ttoff

Both declarations initialize all components to be zero. The array can
have arbitrary many indices (\emph{slots}) and they are counted
starting at {\tt 0}. The are no special operators defined for handling
arrays -- all operations have to be done \emph{by hand}, mostly with
{\tt for} loops running over the indices.

The matrix has exactly two indices which start counting at {\tt 1}! A
matrix can also be constructed by the {\tt mat} command where the
components have to be structured in tuples: e.g.\ {\tt mat
  ((1,2,3),(4,5,6));} constructs a $2\!\times\!3$-matrix that Reduce
displays as

\verbon\begin{verbatim}
[1  2  3]
[       ]
[4  5  6]
\end{verbatim}
\verboff

Reduce offers the following routines to handle matrices:

\tton
tp trace det rank mateigen ({\it matrix});
\ttoff

They calculate the transpose, trace, determinate, rank, and the
eigenvalue equation and eigenvectors of the given {\it matrix},
respectively. Also, matrices can be added, multiplied, and divided. (A
division means multiplication with the inverse.) A matrix can be
constructed via the

\section{Export / import}

\subsection{Storing results}

Storing results is very important for large problems. Reduce offers no
extra command for storing expressions but it is quite easy to write a
little procedure that stores e.g.\ an arbitrary list of names together
with their values. For this we write appropriate assignments into an
extra file with the {\tt nat} switch turned off. Such a procedure could
read

\verbon\begin{verbatim}
procedure store(filename,namelist,valuelist)$ begin
  off nat$
  out filename$
  for each name in namelist do <<
    write name,":=",first valuelist$
    valuelist:=rest valuelist$
  >>$
  write "end"$
  shut filename$
  on nat$
end$
\end{verbatim}
\verboff

With this procedure defined we can execute the lines

\tton
f:=sin(x)\$ g:=cos(x)\$\\
store("FandG",{"f","g"},{f,g})\$\\
clear f,g\$\\
in "FandG"\$\\
f;g;
\ttoff

where, the filename {\tt "FandG"}, the name list {\tt \{"f","g"\}},
and their values {\tt \{f,g\}} is passed to the procedure {\tt store}.
This produces the file {\tt "FandG"} in the current directory reading

\tton
f:=sin(x)\$\\
g:=cos(x)\$\\
end\$
\ttoff

such that the {\tt in} command reassigns their values to {\tt f} and
{\tt g}.

\subsection{TeX}

To export algebraic expression to TeX you need to load the package
{\tt tri}. This packages provides two switches {\tt tex} and {\tt
  texbreak} which causes Reduce to format any output in TeX syntax
({\tt texbreak} also breaks the line in long equations). Example:

\verbon\begin{verbatim}
1: load_package tri;
*** global `metricu!*' cannot become fluid
*** global `indxl!*' cannot become fluid
% TeX-REDUCE-Interface 0.50
% set greek asserted
% set lowercase asserted
% \tolerance 10
% \hsize=150mm

2: depend(f,x);

3: df(f,x);
$$
\frac{\partial f}{
      \partial x}
$$
\end{verbatim}
\verboff

For some TeX-symbols Reduce uses own macros which can be included in
TeX with an {\tt \verb+\+input\{tridefs\}} command, (You can find the
{\tt tridefs.tex} file at \cite{mypage}.) To export all the output of
a Reduce file in one file in TeX format one could add the lines

\verbon\begin{verbatim}
load_package tri$
off msg$
on texbreak$
out "sample.out.tex"$
\end{verbatim}
\verboff

at the beginning of the Reduce file and the following lines at the end

\verbon\begin{verbatim}
shut "sample.out.tex"$
quit$
\end{verbatim}
\verboff

Turning off the {\tt msg} switch prevents errors. For further
information on the {\tt tri} package see\\ 
\verb+http://www.uni-koeln.de/REDUCE/3.6/doc/tri.ps+

\subsection{Fortran}

Exporting to Fortran is very similar to exporting to TeX. Reduce
offers the {\tt fort} switch to display all output in Fortran syntax.
Unfortunately, Reduce can not generate whole Fortran procedures - as
Maple does, e.g..

\subsection{Maple}

Reduce offers, of course, no commands to export expressions and
equations to Maple. However, in order to profit form the merits of
both systems it is advantageous to know how to transport expression
between them. Here we discuss how to export from Reduce to Maple.

The strategy I use is to collect all expression in one list. As first
and last entry I insert {\tt "[null"} and {\tt "null]"}, which are the
braces to form ordered lists in Maple. Then I introduce rules to
replace expressions which Maple does not understand, e.g.\ {\tt let \{
  f=>f(x), df(f,x)=>diff(f(x),x) \};}. After reformulating the list
one switches {\tt nat} off and writes the list into a file. See the
following example to pass a list of two expressions to Maple

\verbon\begin{verbatim}
depend(f,x)$

e1:=int(f,x)+f+x**2$
e2:=df(f,x)+sin(x*f/3)$

maplist:={"[null",e1,e2,"null]"}$

operator f,diff$
let f=f(x),df(f,x)=diff(f(x),x)$
maplist:=maplist$

off nat$
out "sample.map"$
write "maplist:=op(",maplist,");"$
shut "sample.map"$
on nat;
\end{verbatim}
\verboff

which results in the file {\tt sample.map}:

\verbon\begin{verbatim}
maplist:=op({[null,
f(x) + int(f(x),x) + x**2,
diff(f(x),x) + sin((f(x)*x)/3),
null]});$
\end{verbatim}
\verboff

This file can be read by Maple with the command {\tt
  read("sample.map"):}. Although the {\tt \$} sign at the end of the
file will cause an error message, the {\tt maplist} is read correctly
(the LaTeX code for the following line was produced by Maple with the
{\tt latex} command):

\mapon
{\it maplist}:=\left[{\it null},f(x)+\int \!f(x){dx}+{x}^{2},{\frac {d}{dx}}f(x)+\sin(1/3
\,f(x)x),{\it null}\right]
\mapoff

\subsection{Gnuplot}

\begin{figure}[t]\center
\begin{minipage}{8cm}
\includegraphics[scale=0.5]{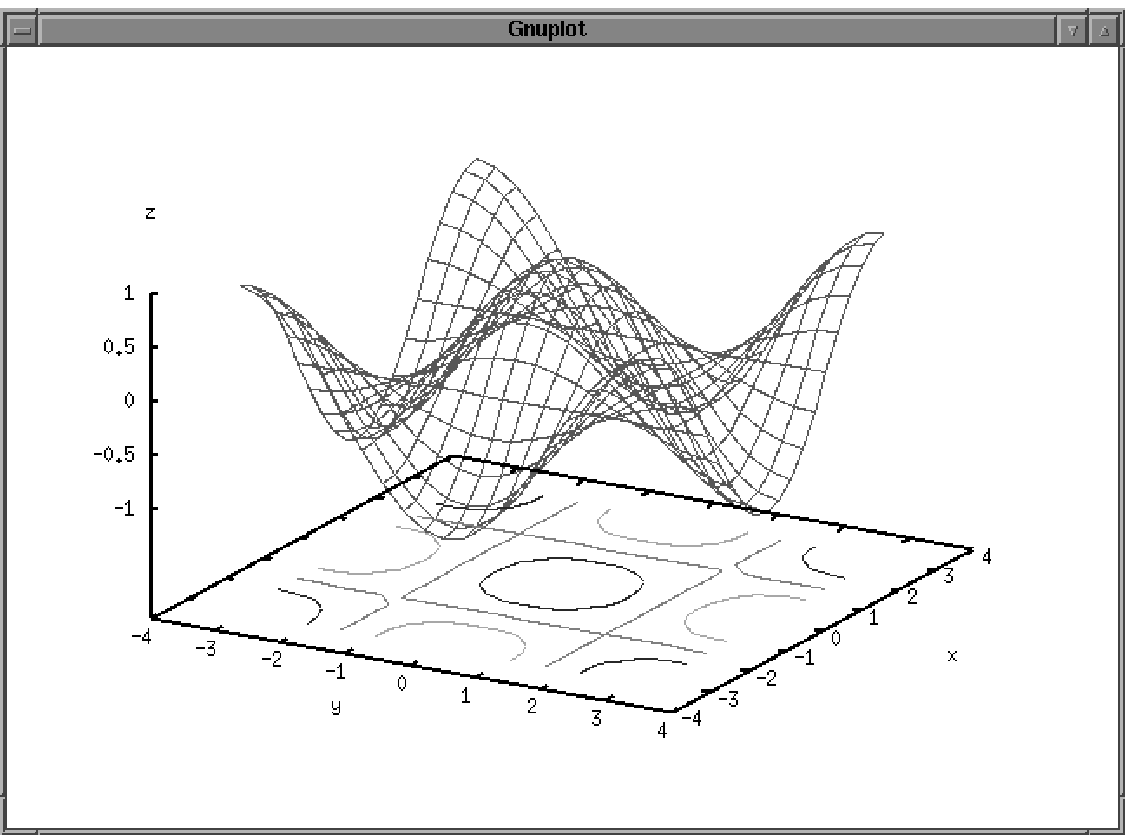}
\end{minipage}
\begin{minipage}{8cm}
\includegraphics[scale=0.25,angle=-90]{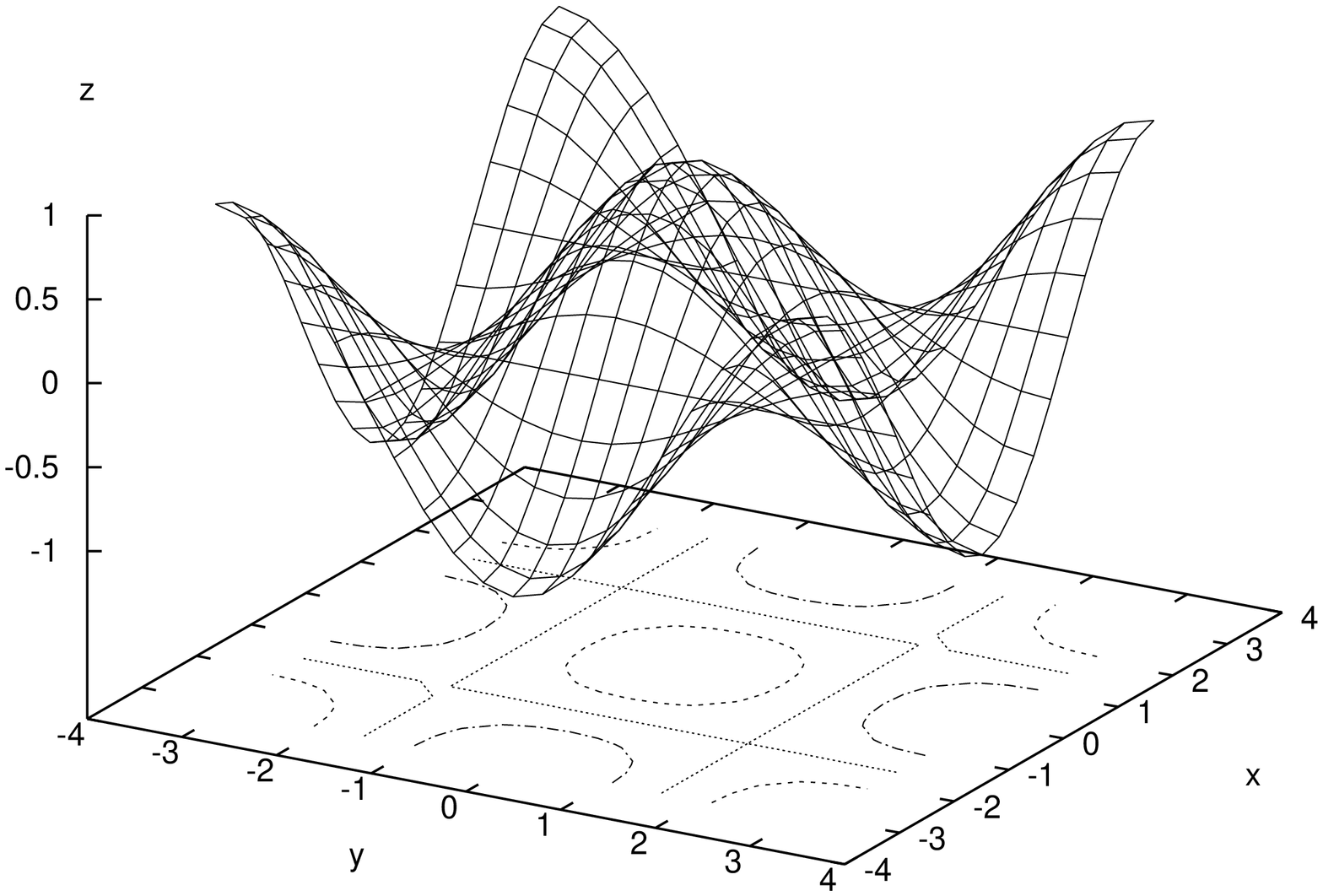}
\end{minipage}
\caption{The gnuplot window and the postscript produced by gnuplot}
\label{Fgnuplot}
\end{figure}

To display functions with gnuplot and to save them as postscripts you
need to load the package {\tt gnuplot}. This package provides the
commands {\tt plot, gnuplot, plotreset}, and {\tt plotshow}. With {\tt
  plot} you will automatically open a gnuplot window that displays
your graph. Example:

\tton
load\_package gnuplot\$\\
plot(cos(x)*cos(y),x=(-pi..pi),y=(-pi..pi),contour);
\ttoff

will open a window that looks like the one in figure \ref{Fgnuplot}

The syntax of this {\tt plot} command is similar to that in gnuplot.
The command

\tton
gnuplot({\it cmd},{\it param1},{\it param2},...);
\ttoff

is supposed to execute all other gnuplot commands (which it does not
always!) The parameters of the gnuplot command {\it com} have to be
separated by commas (not by blanks as in gnuplot). Example:

\tton
gnuplot(set,logscale,y)\$\\
plot(x**2);
\ttoff

will produce a logarithmic plot of $x^2$. Finally, plotting into a file in postscript format is achieved by adding the options {\tt terminal=postscript} and {\tt output={\it filename}} to the plot command. Example:

\tton
plot(cos(x)*cos(y),x=(-pi..pi),y=(-pi..pi),contour,terminal=postscript,output="gnutest.eps");
\ttoff

produces the postscript displayed in figure \ref{Fgnuplot}. For more
information on the {\tt gnuplot} package see\\ 
\verb+http://www.zib.de/Symbolik/reduce/moredocs/gnuplot.pdf+

\chapter{Maple I}

\section{The simplification principle}

Just as Reduce, Maple is an input-output machine. But, as main
difference to Reduce, Maple will not automatically reformulate any
input obeying some rules. Instead, Maple only executes explicit
commands in the input but cites expressions without commands verbally.
Hence, {\bf the simplification performance of Maple is controlled with
  explicit commands only - not with automatic reformulation rules.}
This nature of Maple might stem from its origin: Maple is programmed
in C and the command syntax is strongly influenced by this language.
There exist no rules or switches, Maple becomes active only when a
command or an operator is called.

Maple offers a variety of commands to reformulate expressions. Most
important though is one command, {\tt simplify}, which tries to apply
many different simplification methods to the expression and can be
controlled with some options. For example, Maple will simply cite an
input {\tt (x**2-1)/(x+1);} as

\mapon
\frac{x^2-1}{x+1}
\mapoff

but the answer on {\tt simplify((x**2-1)/(x+1));} is

\mapon
x-1
\mapoff

The general syntax of {\tt simplify} is

\tton
simplify ({\it expr} [,{\it simplification\_methods}] [,assume={\it assumptions}] );
\ttoff

Without specifying the simplification methods, {\tt simplify} applies
all methods. The parameter {\it simplification\_methods} can be the
{\it name} of a user-defined simplification procedure {\tt
  `simplify/{\it name}` ({\it expr})} or any of the build-in methods

\tton 
BesselI, BesselJ, BesselK, BesselY, D, Ei, GAMMA, RootOf,
LambertW, dilog, exp, ln, sqrt, polylog, pochhammer, trig ({\it for trig
functions}), hypergeom ({\it for hypergeometrics}), radical ({\it occurrence of
fractional powers}), power ({\it occurrence of powers}), exp, ln, and atsign
({\it "@" - for operators})
\ttoff

One main problem with Maple's simplification principle is, that {\tt
  simplify} can not handle unknown (e.g.\ package-defined or
user-defined) objects as expressions. For a tensor, e.g., the
simplification command has to be applied on each component separately.
Here, Reduce is clearly more elegant.

\section{The interface}

When you start a graphical version of Maple ('xmaple') you are offered
a \emph{worksheet} displaying the first prompt {\tt > ~}. Behind the
prompt, Maple expects a command consisting out of an statement and a
terminator ({\tt ;} or {\tt :}). Again, the terminator decides whether
Maple's response is displayed ({\tt ;}) or not ({\tt :}). {\bf Note
  that Maple is case sensitive!} All examples displayed in this
chapter are collected in the file {\tt examples.mws} at \cite{mypage}.
The LaTeX code to display the examples here was produced by Maple (see
Import / export).

\footnotesize
\begin{maplegroup}
Simplifying

\begin{mapleinput}
\mapleinline{active}{1d}{(x**2-1)/(x+1);}{%
}
\end{mapleinput}

\mapleresult
\begin{maplelatex}
\mapleinline{inert}{2d}{(x^2-1)/(x+1);}{%
\[
{\displaystyle \frac {x^{2} - 1}{x + 1}} 
\]
}
\end{maplelatex}

\end{maplegroup}
\begin{maplegroup}
\begin{mapleinput}
\mapleinline{active}{1d}{simplify((x**2-1)/(x+1));}{%
}
\end{mapleinput}

\mapleresult
\begin{maplelatex}
\mapleinline{inert}{2d}{x-1;}{%
\[
x - 1
\]
}
\end{maplelatex}

\end{maplegroup}
\begin{maplegroup}
\begin{mapleinput}
\mapleinline{active}{1d}{expand((x+y)**3);}{%
}
\end{mapleinput}

\mapleresult
\begin{maplelatex}
\mapleinline{inert}{2d}{x^3+3*x^2*y+3*x*y^2+y^3;}{%
\[
x^{3} + 3\,x^{2}\,y + 3\,x\,y^{2} + y^{3}
\]
}
\end{maplelatex}

\end{maplegroup}
\begin{maplegroup}
\begin{mapleinput}
\mapleinline{active}{1d}{A:=[(x^2-1)/(x+1),0];simplify(A);}{%
}
\end{mapleinput}

\mapleresult
\begin{maplelatex}
\mapleinline{inert}{2d}{A := [(x^2-1)/(x+1), 0];}{%
\[
A := [{\displaystyle \frac {x^{2} - 1}{x + 1}} , \,0]
\]
}
\end{maplelatex}

\begin{maplelatex}
\mapleinline{inert}{2d}{[x-1, 0];}{%
\[
[x - 1, \,0]
\]
}
\end{maplelatex}

\end{maplegroup}
\begin{maplegroup}
\begin{mapleinput}
\mapleinline{active}{1d}{A:=[0,0];simplify(A);}{%
}
\end{mapleinput}

\mapleresult
\begin{maplelatex}
\mapleinline{inert}{2d}{A := [0, 0];}{%
\[
A := [0, \,0]
\]
}
\end{maplelatex}

\begin{maplelatex}
\mapleinline{inert}{2d}{[0, 0];}{%
\[
[0, \,0]
\]
}
\end{maplelatex}

\end{maplegroup}
\begin{maplegroup}
\begin{mapleinput}
\mapleinline{active}{1d}{A:=array([[(x**2-1)/(x+1),0],[0,0]]); simplify(A);}{%
}
\end{mapleinput}

\mapleresult
\begin{maplelatex}
\mapleinline{inert}{2d}{A := matrix([[(x^2-1)/(x+1), 0], [0, 0]]);}{%
\[
A :=  \left[ 
{\begin{array}{cr}
{\displaystyle \frac {x^{2} - 1}{x + 1}}  & 0 \\ [2ex]
0 & 0
\end{array}}
 \right] 
\]
}
\end{maplelatex}

\begin{maplelatex}
\mapleinline{inert}{2d}{matrix([[x-1, 0], [0, 0]]);}{%
\[
 \left[ 
{\begin{array}{cr}
x - 1 & 0 \\
0 & 0
\end{array}}
 \right] 
\]
}
\end{maplelatex}

\end{maplegroup}
\begin{maplegroup}
\begin{mapleinput}
\mapleinline{active}{1d}{A:=array([[0,0],[0,0]]); simplify(A);}{%
}
\end{mapleinput}

\mapleresult
\begin{maplelatex}
\mapleinline{inert}{2d}{A := matrix([[0, 0], [0, 0]]);}{%
\[
A :=  \left[ 
{\begin{array}{rr}
0 & 0 \\
0 & 0
\end{array}}
 \right] 
\]
}
\end{maplelatex}

\begin{maplelatex}
\mapleinline{inert}{2d}{matrix([[x-1, 0], [0, 0]]);}{%
\[
 \left[ 
{\begin{array}{cr}
x - 1 & 0 \\
0 & 0
\end{array}}
 \right] 
\]
}
\end{maplelatex}

\end{maplegroup}
\normalsize

Note that Maple makes an error in the last line because it does not
handle arrays correctly! The operator {\tt \%} has the value of the
last response of Maple (cf.\ {\tt ws} in Reduce), i.e. in our case
$\frac{x^2-1}{x+1}$.  Similar, {\tt \%\%} and {\tt \%\%\%} represent
the second and third last responses of Maple.

Usually, after typing the terminator one presses ENTER to execute the
command. However, Maple offers to group commands in one so-called
execution groups. To insert a line feed in this group one presses
SHIFT+ENTER. Pressing ENTER somewhere in an execution group then
executes all commands in this group.

\section{User definitions}

\subsection{Names and assignments}

\footnotesize
\begin{maplegroup}
Assignments

\begin{mapleinput}
\mapleinline{active}{1d}{my_name123:=int(f(x),x);}{%
}
\end{mapleinput}

\mapleresult
\begin{maplelatex}
\mapleinline{inert}{2d}{my_name123 := int(f(x),x);}{%
\[
\mathit{my\_name123} := {\displaystyle \int } \mathrm{f}(x)\,dx
\]
}
\end{maplelatex}

\end{maplegroup}
\begin{maplegroup}
\begin{mapleinput}
\mapleinline{active}{1d}{`$& "' *@#!{}`:=int(g(x),x);}{%
}
\end{mapleinput}

\mapleresult
\begin{maplelatex}
\mapleinline{inert}{2d}{`$& "' *@#!{}` := int(g(x),x);}{%
\[
\mathit{\$\&\ ``^{\prime }\ *@\#!} := {\displaystyle \int } 
\mathrm{g}(x)\,dx
\]
}
\end{maplelatex}

\end{maplegroup}
\begin{maplegroup}
\begin{mapleinput}
\mapleinline{active}{1d}{my_name123:='my_name123';
`$& "' *@#!{}` := '`$& "' *@#!{}`';}{%
}
\end{mapleinput}

\mapleresult
\begin{maplelatex}
\mapleinline{inert}{2d}{my_name123 := 'my_name123';}{%
\[
\mathit{my\_name123} := \mathit{my\_name123}
\]
}
\end{maplelatex}

\begin{maplelatex}
\mapleinline{inert}{2d}{`$& "' *@#!{}` := '`$& "' *@#!{}`';}{%
\[
\mathit{\$\&\ ``^{\prime }\ *@\#!} := \mathit{\$\&\ ``^{\prime }
\ *@\#!}
\]
}
\end{maplelatex}

\end{maplegroup}
\begin{maplegroup}
\begin{mapleinput}
\mapleinline{active}{1d}{i:=0:j:=2:k:=1: A.i.j.k;}{%
}
\end{mapleinput}

\mapleresult
\begin{maplelatex}
\mapleinline{inert}{2d}{A021;}{%
\[
\mathit{A021}
\]
}
\end{maplelatex}

\end{maplegroup}
\normalsize

As in Reduce any string (which is not reserved or a keyword, see table
\ref{TreservedMaple}) may be used as a \emph{name}. If you want some
unusual string to be a name, the string has to be embraced by
back(!)-quotes, e.g. {\tt `123 string`} can be used as a name. The
{\tt .} sign allows to append numbers to a name.

The assignment operator is {\tt {\it name}:={\it expr}}. As in Reduce,
the lhs name is identified with the rhs expression. There are three
methods to unassign a name {\tt x}:\\ -- First with {\tt x:='x';}. The
embracing by single quotes(!) means an \emph{unevaluation} of the
enclosed string. Hence, {\tt x:='x';} means an assignment of the
unevaluated string {\tt x} back to {\tt x}.\\ -- Second with {\tt
  unassign('x');}.  The {\tt unassign} command also needs an unevalued
string as parameter.\\ -- Third with {\tt restart;} which, however,
unassigns \emph{all} names. This methods is most secure to free the
memory resource maple allocated.

\begin{table}[t]\center
\begin{tabular}{p{17ex}|p{69ex}}
constants & {\tt Pi I infinity gamma Catalan true false} \spc\\
internal variables & {\tt Digits Order constants libname printlevel lasterror status} \spc\\
keywords & {\tt and by do done elif else end fi for from if in intersect
local minus mod not od option options or proc quit read save stop then
to union while }
\end{tabular}
\caption{Reserved names in Maple.}
\label{TreservedMaple}
\end{table}

As in Reduce it is important to distinguish between the expression
assigned to a name and the value that Maple displays when evaluating
the name. The following example shows the importance of the
reassignment {\tt a:=a:}. See also section \ref{Reducedefs}.

\footnotesize
\begin{maplegroup}
\begin{mapleinput}
\mapleinline{active}{1d}{a:=b: a;}{%
}
\end{mapleinput}

\mapleresult
\begin{maplelatex}
\mapleinline{inert}{2d}{b;}{%
\[
b
\]
}
\end{maplelatex}

\end{maplegroup}
\begin{maplegroup}
\begin{mapleinput}
\mapleinline{active}{1d}{b:=1: a;}{%
}
\end{mapleinput}

\mapleresult
\begin{maplelatex}
\mapleinline{inert}{2d}{1;}{%
\[
1
\]
}
\end{maplelatex}

\end{maplegroup}
\begin{maplegroup}
\begin{mapleinput}
\mapleinline{active}{1d}{a:=a:b:=2: a;}{%
}
\end{mapleinput}

\mapleresult
\begin{maplelatex}
\mapleinline{inert}{2d}{1;}{%
\[
1
\]
}
\end{maplelatex}

\end{maplegroup}
\begin{maplegroup}
\begin{mapleinput}
\mapleinline{active}{1d}{a:='a';
b:='b';}{%
}
\end{mapleinput}

\mapleresult
\begin{maplelatex}
\mapleinline{inert}{2d}{a := 'a';}{%
\[
a := a
\]
}
\end{maplelatex}

\begin{maplelatex}
\mapleinline{inert}{2d}{b := 'b';}{%
\[
b := b
\]
}
\end{maplelatex}

\end{maplegroup}
\normalsize

\subsection{Aliases}

\footnotesize
\begin{maplegroup}
Aliases

\begin{mapleinput}
\mapleinline{active}{1d}{alias(I='I'):macro(I='I'): 2*I; sqrt(-4);}{%
}
\end{mapleinput}

\mapleresult
\begin{maplelatex}
\mapleinline{inert}{2d}{2*I;}{%
\[
2\,I
\]
}
\end{maplelatex}

\begin{maplelatex}
\mapleinline{inert}{2d}{2*sqrt(-1);}{%
\[
2\,\sqrt{-1}
\]
}
\end{maplelatex}

\end{maplegroup}
\begin{maplegroup}
\begin{mapleinput}
\mapleinline{active}{1d}{macro(I=sqrt(-1)): 2*I; sqrt(-4);}{%
}
\end{mapleinput}

\mapleresult
\begin{maplelatex}
\mapleinline{inert}{2d}{2*sqrt(-1);}{%
\[
2\,\sqrt{-1}
\]
}
\end{maplelatex}

\begin{maplelatex}
\mapleinline{inert}{2d}{2*sqrt(-1);}{%
\[
2\,\sqrt{-1}
\]
}
\end{maplelatex}

\end{maplegroup}
\begin{maplegroup}
\begin{mapleinput}
\mapleinline{active}{1d}{macro(I='I'): 2*I; sqrt(-4);}{%
}
\end{mapleinput}

\mapleresult
\begin{maplelatex}
\mapleinline{inert}{2d}{2*I;}{%
\[
2\,I
\]
}
\end{maplelatex}

\begin{maplelatex}
\mapleinline{inert}{2d}{2*sqrt(-1);}{%
\[
2\,\sqrt{-1}
\]
}
\end{maplelatex}

\end{maplegroup}
\begin{maplegroup}
\begin{mapleinput}
\mapleinline{active}{1d}{alias(I=sqrt(-1)): 2*I; sqrt(-4);}{%
}
\end{mapleinput}

\mapleresult
\begin{maplelatex}
\mapleinline{inert}{2d}{2*I;}{%
\[
2\,I
\]
}
\end{maplelatex}

\begin{maplelatex}
\mapleinline{inert}{2d}{2*I;}{%
\[
2\,I
\]
}
\end{maplelatex}

\end{maplegroup}
\begin{maplegroup}
\begin{mapleinput}
\mapleinline{active}{1d}{alias(I='I'): 2*I; sqrt(-4);}{%
}
\end{mapleinput}

\mapleresult
\begin{maplelatex}
\mapleinline{inert}{2d}{2*I;}{%
\[
2\,I
\]
}
\end{maplelatex}

\begin{maplelatex}
\mapleinline{inert}{2d}{2*sqrt(-1);}{%
\[
2\,\sqrt{-1}
\]
}
\end{maplelatex}

\end{maplegroup}
\normalsize

Maple has two possibilities to define aliases which are quite useful
(in contrast to the {\tt define} command in Reduce). The first is {\tt
  macro} and the second is {\tt alias}. They have the same syntax but
they have different meaning:

The {\tt macro({\it name}={\it expr});} command does \emph{not}
evaluate the expression {\it expr} before assigning it to the macro's
{\it name}. Whenever this {\it name} will appear in a future expression
the \emph{first} thing Maple does is to replace it by the associated
value {\it expr}. However, Maple will not replace {\it expr} by the
macro {\it name} in outputs. The point of a macro is to introduce
shorthand notations \emph{for the input}.

The {\tt alias({\it name}={\it expr});} command \emph{evaluates} the
expression {\it expr} before assigning it to the {\it name}. The value
of {\it expr} cannot be a number. The effect of an alias is first,
that Maple replaces all aliases by its value before evaluating
expressions (as it does with macros), but second, that Maple also
replaces expressions in the output by its alias if these exists one.
So, the main point of an alias is introduce abbreviations in the input
\emph{and output}. The best example: The complex unit {\tt I} is
defined in Maple via: {\tt alias(I=sqrt(-1));}

Macros and aliases can be changed and can be unassigned by assigning
them back to their names: {\tt macro({\it name}={\it name});} and {\tt
  alias({\it name}={\it name});}

\subsection{Substitutions and assumptions}

\footnotesize
\begin{maplegroup}
Substitutions

\begin{mapleinput}
\mapleinline{active}{1d}{f:=cos(x): x:=0: f; x:='x':}{%
}
\end{mapleinput}

\mapleresult
\begin{maplelatex}
\mapleinline{inert}{2d}{1;}{%
\[
1
\]
}
\end{maplelatex}

\end{maplegroup}
\begin{maplegroup}
\begin{mapleinput}
\mapleinline{active}{1d}{f:=cos(x): simplify(subs(x=0,f));}{%
}
\end{mapleinput}

\mapleresult
\begin{maplelatex}
\mapleinline{inert}{2d}{1;}{%
\[
1
\]
}
\end{maplelatex}

\end{maplegroup}
\begin{maplegroup}
\begin{mapleinput}
\mapleinline{active}{1d}{f:='f':}{%
}
\end{mapleinput}

\end{maplegroup}
\normalsize

As in Reduce, one way to substitute identities in expressions is to
make a local definition that represents the identity and let Maple
reevaluate the expression. After this, {\tt f} is still assigned to
{\tt cos(x)} but was evaluated to be 1. However, one should clear the
assignment for {\tt x} again with {\tt x:='x':}. A more elegant way to
substitute is the {\tt subs} command the syntax of which is:

\tton
subs({\it eqn\_list},{\it expr});
\ttoff

Eventually, one has to {\tt simplify} again to take the substitution
into account.

Maple has one quite surprising and useful tool that Reduce lacks. {\bf
  Maple allows to make assumptions on unassigned names!} With this,
e.g., Maple simplifies $\sqrt{x^2}$ to $x$ if one assumed $x>0$
before:

\footnotesize
\begin{maplegroup}
Assumptions

\begin{mapleinput}
\mapleinline{active}{1d}{sqrt(x^2);}{%
}
\end{mapleinput}

\mapleresult
\begin{maplelatex}
\mapleinline{inert}{2d}{sqrt(x^2);}{%
\[
\sqrt{x^{2}}
\]
}
\end{maplelatex}

\end{maplegroup}
\begin{maplegroup}
\begin{mapleinput}
\mapleinline{active}{1d}{assume(x>0): about(x);}{%
}
\end{mapleinput}

\mapleresult
\begin{maplettyout}
Originally x, renamed x~:
\end{maplettyout}

\begin{maplettyout}
  is assumed to be: RealRange(Open(0),infinity)
\end{maplettyout}

\emptyline
\end{maplegroup}
\begin{maplegroup}
\begin{mapleinput}
\mapleinline{active}{1d}{sqrt(x^2);}{%
}
\end{mapleinput}

\mapleresult
\begin{maplelatex}
\mapleinline{inert}{2d}{x;}{%
\[
\mathit{x\symbol{126}}
\]
}
\end{maplelatex}

\end{maplegroup}
\begin{maplegroup}
\begin{mapleinput}
\mapleinline{active}{1d}{x := 'x';}{%
}
\end{mapleinput}

\mapleresult
\begin{maplelatex}
\mapleinline{inert}{2d}{x := 'x';}{%
\[
x := x
\]
}
\end{maplelatex}

\end{maplegroup}
\begin{maplegroup}
\begin{mapleinput}
\mapleinline{active}{1d}{sqrt(x^2);}{%
}
\end{mapleinput}

\mapleresult
\begin{maplelatex}
\mapleinline{inert}{2d}{sqrt(x^2);}{%
\[
\sqrt{x^{2}}
\]
}
\end{maplelatex}

\end{maplegroup}
\normalsize

The syntax is {\tt assume({\it properties});} The twiddle \~{} behind
the name {\tt x} indicates that there are properties assumed for {\tt
  x}. You can ask for properties of names with {\tt about} and {\tt
  is} and add more properties with {\tt additionally}.  Properties are
deleted with the reassignment {\tt x:='x';}

\subsection{Operators}

\footnotesize
\begin{maplegroup}
Operators

\begin{mapleinput}
\mapleinline{active}{1d}{define(log2,log2(2)=1):
log2(2)*log2(3);}{%
}
\end{mapleinput}

\mapleresult
\begin{maplelatex}
\mapleinline{inert}{2d}{log2(3);}{%
\[
\mathrm{log2}(3)
\]
}
\end{maplelatex}

\end{maplegroup}
\begin{maplegroup}
\begin{mapleinput}
\mapleinline{active}{1d}{define(fac,fac(0)=1,fac(n::posint)=n*fac(n-1)):
fac(5);}{%
}
\end{mapleinput}

\mapleresult
\begin{maplelatex}
\mapleinline{inert}{2d}{120;}{%
\[
120
\]
}
\end{maplelatex}

\end{maplegroup}
\begin{maplegroup}
\begin{mapleinput}
\mapleinline{active}{1d}{unassign(log2,fac):}{%
}
\end{mapleinput}

\end{maplegroup}
\normalsize

Maple offers the {\tt define} commands to declare new operators.
The general syntax is

\tton
define ({\it name},{\it properties});\\
definemore ({\it name},{\it properties});
\ttoff

Here, {\it properties} are equations (rules) or any of the following
key properties: {\tt linear, multilinear, orderless, flat ({\it means
    associative})}. Note that it is possible to specify the type (see
sections \ref{Mapleproc} and \ref{Mapletypes}) of a parameter (e.g.\ 
{\tt n::posint} to allow only positive integers for the parameter {\tt
  n}). The command {\tt definemore} adds more properties to an already
defined operator.

\subsection{Functionals}

\footnotesize
\begin{maplegroup}
\begin{mapleinput}
\mapleinline{active}{1d}{f(x):=cos(x): f(x); f(0); f(y);}{%
}
\end{mapleinput}

\mapleresult
\begin{maplelatex}
\mapleinline{inert}{2d}{cos(x);}{%
\[
\mathrm{cos}(x)
\]
}
\end{maplelatex}

\begin{maplelatex}
\mapleinline{inert}{2d}{f(0);}{%
\[
\mathrm{f}(0)
\]
}
\end{maplelatex}

\begin{maplelatex}
\mapleinline{inert}{2d}{f(y);}{%
\[
\mathrm{f}(y)
\]
}
\end{maplelatex}

\end{maplegroup}
\begin{maplegroup}
\begin{mapleinput}
\mapleinline{active}{1d}{f(x):='f(x)':}{%
}
\end{mapleinput}

\end{maplegroup}
\begin{maplegroup}
Functionals

\begin{mapleinput}
\mapleinline{active}{1d}{g := x -> cos(x); g(x); g(0); g(y); D(g);}{%
}
\end{mapleinput}

\mapleresult
\begin{maplelatex}
\mapleinline{inert}{2d}{g := cos;}{%
\[
g := \mathrm{cos}
\]
}
\end{maplelatex}

\begin{maplelatex}
\mapleinline{inert}{2d}{cos(x);}{%
\[
\mathrm{cos}(x)
\]
}
\end{maplelatex}

\begin{maplelatex}
\mapleinline{inert}{2d}{1;}{%
\[
1
\]
}
\end{maplelatex}

\begin{maplelatex}
\mapleinline{inert}{2d}{cos(y);}{%
\[
\mathrm{cos}(y)
\]
}
\end{maplelatex}

\begin{maplelatex}
\mapleinline{inert}{2d}{-sin;}{%
\[
 - \mathrm{sin}
\]
}
\end{maplelatex}

\end{maplegroup}
\begin{maplegroup}
\begin{mapleinput}
\mapleinline{active}{1d}{h := (x,y) -> sin(x)*sin(y); D[1](h); D[2](h);}{%
}
\end{mapleinput}

\mapleresult
\begin{maplelatex}
\mapleinline{inert}{2d}{h := proc (x, y) options operator, arrow; sin(x)*sin(y) end;}{%
\[
h := (x, \,y)\rightarrow \mathrm{sin}(x)\,\mathrm{sin}(y)
\]
}
\end{maplelatex}

\begin{maplelatex}
\mapleinline{inert}{2d}{proc (x, y) options operator, arrow; cos(x)*sin(y) end;}{%
\[
(x, \,y)\rightarrow \mathrm{cos}(x)\,\mathrm{sin}(y)
\]
}
\end{maplelatex}

\begin{maplelatex}
\mapleinline{inert}{2d}{proc (x, y) options operator, arrow; sin(x)*cos(y) end;}{%
\[
(x, \,y)\rightarrow \mathrm{sin}(x)\,\mathrm{cos}(y)
\]
}
\end{maplelatex}

\end{maplegroup}
\begin{maplegroup}
\begin{mapleinput}
\mapleinline{active}{1d}{k := h @ (x -> (x,exp(x))); k(x);}{%
}
\end{mapleinput}

\mapleresult
\begin{maplelatex}
\mapleinline{inert}{2d}{k := `@`(h,proc (x) options operator, arrow; x, exp(x) end);}{%
\[
k := h\mathrm{@}(x\rightarrow (x, \,e^{x}))
\]
}
\end{maplelatex}

\begin{maplelatex}
\mapleinline{inert}{2d}{sin(x)*sin(exp(x));}{%
\[
\mathrm{sin}(x)\,\mathrm{sin}(e^{x})
\]
}
\end{maplelatex}

\end{maplegroup}
\begin{maplegroup}
\begin{mapleinput}
\mapleinline{active}{1d}{sin@exp; (sin@exp)(x); D(sin@exp); (D(sin@exp)) (x);}{%
}
\end{mapleinput}

\mapleresult
\begin{maplelatex}
\mapleinline{inert}{2d}{`@`(sin,exp);}{%
\[
\mathrm{sin}\mathrm{@}\mathrm{exp}
\]
}
\end{maplelatex}

\begin{maplelatex}
\mapleinline{inert}{2d}{sin(exp(x));}{%
\[
\mathrm{sin}(e^{x})
\]
}
\end{maplelatex}

\begin{maplelatex}
\mapleinline{inert}{2d}{`@`(cos,exp)*exp;}{%
\[
\mathrm{cos}\mathrm{@}\mathrm{exp}\,\mathrm{exp}
\]
}
\end{maplelatex}

\begin{maplelatex}
\mapleinline{inert}{2d}{cos(exp(x))*exp(x);}{%
\[
\mathrm{cos}(e^{x})\,e^{x}
\]
}
\end{maplelatex}

\end{maplegroup}
\begin{maplegroup}
\begin{mapleinput}
\mapleinline{active}{1d}{sin@@3; (sin@@3)(x);}{%
}
\end{mapleinput}

\mapleresult
\begin{maplelatex}
\mapleinline{inert}{2d}{`@@`(sin,3);}{%
\[
\mathrm{sin}^{(3)}
\]
}
\end{maplelatex}

\begin{maplelatex}
\mapleinline{inert}{2d}{`@@`(sin,3)(x);}{%
\[
(\mathrm{sin}^{(3)})(x)
\]
}
\end{maplelatex}

\end{maplegroup}
\begin{maplegroup}
\begin{mapleinput}
\mapleinline{active}{1d}{g:='g': f:='f': h:='h': k:='k':}{%
}
\end{mapleinput}

\end{maplegroup}
\normalsize

\section{Solving}

Maple offers the {\tt solve} command for solving equations. The syntax is

\tton
solve ({\it eqn\_list},{\it name\_list});
\ttoff

For solving differential equations, use {\tt dsolve}. For purely
floating-point solutions use {\tt fsolve}. Use {\tt isolve} to solve
for integer solutions, {\tt msolve} to solve modulo a prime; {\tt
  rsolve} for recurrences, and {\tt linalg[linsolve]} to solve matrix
equations.

\footnotesize
\begin{maplegroup}
Solving

\begin{mapleinput}
\mapleinline{active}{1d}{solve(x^4-5*x^2+6*x=2,x);}{%
}
\end{mapleinput}

\mapleresult
\begin{maplelatex}
\mapleinline{inert}{2d}{-1+sqrt(3), -1-sqrt(3), 1, 1;}{%
\[
 - 1 + \sqrt{3}, \, - 1 - \sqrt{3}, \,1, \,1
\]
}
\end{maplelatex}

\end{maplegroup}
\begin{maplegroup}
\begin{mapleinput}
\mapleinline{active}{1d}{diff(y(x),x)-y(x)^2+y(x)*sin(x)-cos(x); dsolve(\%);}{%
}
\end{mapleinput}

\mapleresult
\begin{maplelatex}
\mapleinline{inert}{2d}{diff(y(x),x)-y(x)^2+y(x)*sin(x)-cos(x);}{%
\[
({\frac {\partial }{\partial x}}\,\mathrm{y}(x)) - \mathrm{y}(x)
^{2} + \mathrm{y}(x)\,\mathrm{sin}(x) - \mathrm{cos}(x)
\]
}
\end{maplelatex}

\begin{maplelatex}
\mapleinline{inert}{2d}{y(x) = -exp(-cos(x))/(_C1+Int(exp(-cos(x)),x))+sin(x);}{%
\[
\mathrm{y}(x)= - {\displaystyle \frac {e^{( - \mathrm{cos}(x))}}{
\mathit{\_C1} + {\displaystyle \int } e^{( - \mathrm{cos}(x))}\,d
x}}  + \mathrm{sin}(x)
\]
}
\end{maplelatex}

\end{maplegroup}
\begin{maplegroup}
\begin{mapleinput}
\mapleinline{active}{1d}{diff(x(t),t)=y(t); diff(y(t),t)=x(t)+y(t); x(0)=2,y(0)=1;}{%
}
\end{mapleinput}

\mapleresult
\begin{maplelatex}
\mapleinline{inert}{2d}{diff(x(t),t) = y(t);}{%
\[
{\frac {\partial }{\partial t}}\,\mathrm{x}(t)=\mathrm{y}(t)
\]
}
\end{maplelatex}

\begin{maplelatex}
\mapleinline{inert}{2d}{diff(y(t),t) = x(t)+y(t);}{%
\[
{\frac {\partial }{\partial t}}\,\mathrm{y}(t)=\mathrm{x}(t) + 
\mathrm{y}(t)
\]
}
\end{maplelatex}

\begin{maplelatex}
\mapleinline{inert}{2d}{x(0) = 2, y(0) = 1;}{%
\[
\mathrm{x}(0)=2, \,\mathrm{y}(0)=1
\]
}
\end{maplelatex}

\end{maplegroup}
\begin{maplegroup}
\begin{mapleinput}
\mapleinline{active}{1d}{res := dsolve(\{\%\%\%,\%\%,\%\},\{x(t),y(t)\}, type=numeric, output=procedurelist);}{%
}
\end{mapleinput}

\mapleresult
\begin{maplelatex}
\mapleinline{inert}{2d}{res := proc (rkf45_x) local i, rkf45_s, outpoint, r1, r2; global
loc_control, loc_y0, loc_y1; option `Copyright (c) 1993 by the
University of Waterloo. All rights reserved.`; outpoint :=
evalf(rkf45_x); if abs(-outpoint) < abs(loc_control[2]-outpoint) or
not member(loc_control[6],\{-1, 1, -2, 2, -1., -2., 2., 1.\}) then
loc_control := copy(array(1 ..
33,[(24)=0,(7)=.1e-8,(10)=0,(13)=0,(15)=0,(11)=0,(33)=0,(25)=0,(30)=0,
(3)=0,(28)=0,(8)=30000,(9)=1000,(18)=0,(21)=1.,(22)=0,(2)=0,(17)=0,(20
)=2.,(26)=0,(27)=0,(1)=2,(29)=0,(14)=0,(32)=0,(31)=0,(23)=0,(16)=0,(5)
=.1e-7,(6)=1,(12)=0,(19)=0,(4)=.1e-7])); loc_y0 := copy(array(1 ..
2,[(2)=1.,(1)=2.])); loc_y1 := copy(array(1 .. 2,[])) fi; if
abs(-outpoint) <> 0 then loc_control[3] := outpoint; if Digits <=
evalhf(Digits) then rkf45_s :=
traperror(evalhf(`dsolve/numeric_solnall_rkf45`(loc_F,var(loc_control)
,var(loc_y0),var(loc_y1),var(loc_F1),var(loc_F2),var(loc_F3),var(loc_F
4),var(loc_F5),var(loc_work)))); if rkf45_s = lasterror then r1 :=
searchtext('evalhf',convert(op(1,[rkf45_s]),name)); r2 :=
searchtext('hardware',convert(op(1,[rkf45_s]),name)); if r1 <> 0 or r2
<> 0 then
`dsolve/numeric_solnall_rkf45`(loc_F,loc_control,loc_y0,loc_y1,loc_F1,
loc_F2,loc_F3,loc_F4,loc_F5,loc_work) else ERROR(rkf45_s) fi fi else
`dsolve/numeric_solnall_rkf45`(loc_F,loc_control,loc_y0,loc_y1,loc_F1,
loc_F2,loc_F3,loc_F4,loc_F5,loc_work) fi fi; [t = rkf45_x,
seq(ord[i+1] = loc_y0[i],i = 1 .. 2)] end;}{%
\[
\mathit{res} := \textbf{proc} (\mathit{rkf45\_x})\,\ldots \,
\textbf{end} 
\]
}
\end{maplelatex}

\end{maplegroup}
\begin{maplegroup}
\begin{mapleinput}
\mapleinline{active}{1d}{res(1);}{%
}
\end{mapleinput}

\mapleresult
\begin{maplelatex}
\mapleinline{inert}{2d}{[t = 1, x(t) = 5.582168689244844, y(t) = 7.826891137110794];}{%
\[
[t=1, \,\mathrm{x}(t)=5.582168689244844, \,\mathrm{y}(t)=
7.826891137110794]
\]
}
\end{maplelatex}

\end{maplegroup}
\normalsize

We see how to plot solutions (esp. of {\tt dsolve}) in chapter 4.

\section{Commands and references}

Maple has many, many command - a lot more than Reduce. Here, I only
tried to collect the most basic ones in table \ref{TMaplecoms}. Other
than Reduce, Maple offers complete commands for almost any standard
problem instead of providing only elementary commands that suffice to
write procedures. Many of them might provide quick help.
Unfortunately, this means a vast number of commands which can hardly
be overviewed. Most of these one will never use anyway because they
are too specialized.

\begin{table}[t]\center
\begin{tabular}{p{15ex}|p{74ex}}

system control & {\tt restart \% \%\% \%\%\% quit;~~ \# {\it
    comment};~~ with ({\it package});\newline readlib ({\it
    command\_name});~~ interface print[f]}\spc\\ 

assignment & {\tt {\it name} := {\it expr};~~ alias macro ({\it
    name}={\it expr});~~ assume ({\it properties});}\spc\\ 

unassignment & {\tt unassign ({\it name});~~ x:='x';}\spc\\
 
substitution & {\tt subs asubs ({\it eqn\_list},{\it expr});~~ subsop
  ({\it eqn\_list},{\it expr});}\spc\\ 

elementary & {\tt {\it expr} + - * / ** {\it expr};~~ factorial ({\it
    integer});~~ {\it expr} mod {\it integer};}\spc\\ 

logical & {\tt {\it expr} < <= <> = > >= {\it expr};~~ {\it boolean}
  and not or {\it boolean};}\spc\\ 

composition & {\tt {\it functional} @ {\it functional};~~ {\it
    functional} @@ {\it power};} \spc\\ 

operators & {\tt define ({\it name},{\it properties});}\spc\\

selection & {\tt op ([{\it operand\_number},]{\it expr});~~ denom
  numer ({\it expr});\newline lhs rhs ({\it eqn}); collect
  [t]coeff[s]}\spc\\ 

sets/lists & {\tt \{ {\it set\_elements} \};~~ [ {\it list\_elements}
  ];~~ seq ({\it expr},{\it range});\newline {\it expr} \$ {\it range};~~
  {\it set} intersect minus union {\it set};\newline member map select
  }\spc\\ 

arrays & {\tt array ({\it ranges},{\it component\_list});~~ evalm ({\it matrix\_expression});~~ \&* \&*()}\spc\\

loops & {\tt (for from by [to|while] do od) (for in do od) (while do
  od) break next}\spc\\ 
 
conditions & {\tt (if then elif else fi)}\spc\\ 

procedures & {\tt (proc local global options end)\newline ERROR WARNING
  ({\it message});~~ RETURN ({\it exprs});}\spc\\ 

input output & {\tt latex fortran C ({\it expr});~~ writeto appendto
  ("{\it filename}");\newline save {\it names},"{\it filename}";~~
  read "{\it filename}";\newline readdata("{\it filename}" [,{\it
    format}]);~~ writedata("{\it filename}",{\it list});\newline fopen
  fprintf fscanf fclose}\spc\\ 

plotting & {\tt plot display odeplot textplot plotsetup
  interface}\spc\\ 

evaluation & {\tt E/eval eval[a|gf|m|b|c|hf|f|r] value allvalues {\it expr};}\spc\\ 

solving & {\tt solve [i|m|r|f|d]solve ({\it eqn\_list},{\it name\_list});~~ testeq ({\it eqn});}\spc\\ 

handing types & {\tt whattype ({\it expr});~~ type convert ({\it expr},{\it type});}\spc\\ 
\end{tabular}
\caption{
  Elementary commands in Maple. We do not display the syntax of all
  commands and the innumerous functional operators {\tt sin} etc. For
  explanations see table \ref{TReducecoms}. The slash in {\tt E/eval}
  means that there exist both commands {\tt Eval} and {\tt eval}, the
  first of which is the inert version (non-executing) whereas the
  second tries to execute the operation. The construction {\tt
    eval[a|gf|m|..]} means that there exists {\tt evala evalgf evalm}
  etc.}
\label{TMaplecoms}
\end{table}

Because of these innumerable commands Maple offers the most efficient
way to work with Maple is by using the build-in help browser. This
browser is opened via the programs menu or, more efficiently, by
positioning the cursor on a word and pressing 'Crtl+F1'. To begin
with, it is useful to have a look on the 'Index...' topic (you find in
the upper left window). There you find lists of expressions,
functions, packages, etc. Alternatively, the 'Topic search...' in the
'Help' menu it useful to search for commands etc. alphabetically.

There is also online help in the internet available at
\cite{maplepage}. This page is the producers address. You can find
more addresses at \cite{mypage}.

There is also help available with in the non-graphical interface. The
command {\tt ?{\it subject}} will display help and examples for the
specified subject. Also the commands {\tt index info usage example
  related ({\it subject});} provide help on the subject.

\footnotesize
\begin{maplegroup}
Integration / differentiation /dependencies:

\begin{mapleinput}
\mapleinline{active}{1d}{diff(cos(5*x),x);
int(sin(x)**2,x);}{%
}
\end{mapleinput}

\mapleresult
\begin{maplelatex}
\mapleinline{inert}{2d}{-5*sin(5*x);}{%
\[
 - 5\,\mathrm{sin}(5\,x)
\]
}
\end{maplelatex}

\begin{maplelatex}
\mapleinline{inert}{2d}{-1/2*cos(x)*sin(x)+1/2*x;}{%
\[
 - {\displaystyle \frac {1}{2}} \,\mathrm{cos}(x)\,\mathrm{sin}(x
) + {\displaystyle \frac {1}{2}} \,x
\]
}
\end{maplelatex}

\end{maplegroup}
\begin{maplegroup}
\begin{mapleinput}
\mapleinline{active}{1d}{diff(f(x),x); diff(f,x); int(f(x),x); int(f,x);}{%
}
\end{mapleinput}

\mapleresult
\begin{maplelatex}
\mapleinline{inert}{2d}{diff(cos(x)(x),x);}{%
\[
{\frac {\partial }{\partial x}}\,\mathrm{cos}(x)(x)
\]
}
\end{maplelatex}

\begin{maplelatex}
\mapleinline{inert}{2d}{-sin(x);}{%
\[
 - \mathrm{sin}(x)
\]
}
\end{maplelatex}

\begin{maplelatex}
\mapleinline{inert}{2d}{int(cos(x)(x),x);}{%
\[
{\displaystyle \int } \mathrm{cos}(x)(x)\,dx
\]
}
\end{maplelatex}

\begin{maplelatex}
\mapleinline{inert}{2d}{sin(x);}{%
\[
\mathrm{sin}(x)
\]
}
\end{maplelatex}

\end{maplegroup}
\begin{maplegroup}
\begin{mapleinput}
\mapleinline{active}{1d}{f(x):=cos(x): f(x); f(0); f(y);}{%
}
\end{mapleinput}

\mapleresult
\begin{maplelatex}
\mapleinline{inert}{2d}{cos(x);}{%
\[
\mathrm{cos}(x)
\]
}
\end{maplelatex}

\begin{maplelatex}
\mapleinline{inert}{2d}{f(0);}{%
\[
\mathrm{f}(0)
\]
}
\end{maplelatex}

\begin{maplelatex}
\mapleinline{inert}{2d}{f(y);}{%
\[
\mathrm{f}(y)
\]
}
\end{maplelatex}

\end{maplegroup}
\begin{maplegroup}
\begin{mapleinput}
\mapleinline{active}{1d}{f(x):='f(x)':}{%
}
\end{mapleinput}

\end{maplegroup}
\begin{maplegroup}
Procedures

\begin{mapleinput}
\mapleinline{active}{1d}{my_factorial:=proc(k)
  local x,l: x:=1:
  for l from 1 to k do x:=x*l: od:
end;}{%
}
\end{mapleinput}

\mapleresult
\begin{maplelatex}
\mapleinline{inert}{2d}{my_factorial := proc (k) local x, l; x := 1; for l to k do x := x*l
od end;}{%
\[
\mathit{my\_factorial} := \textbf{proc} (k)\,\textbf{local} \,x, 
\,l;\,x := 1\,;\,\textbf{for} \,l\,\textbf{to} \,k\,\textbf{do} 
\,x := x \times l\,\textbf{od} \,\textbf{end} 
\]
}
\end{maplelatex}

\end{maplegroup}
\begin{maplegroup}
\begin{mapleinput}
\mapleinline{active}{1d}{my_factorial(16);}{%
}
\end{mapleinput}

\mapleresult
\begin{maplelatex}
\mapleinline{inert}{2d}{20922789888000;}{%
\[
20922789888000
\]
}
\end{maplelatex}

\end{maplegroup}
\begin{maplegroup}
\begin{mapleinput}
\mapleinline{active}{1d}{unassign(my_factorial):}{%
}
\end{mapleinput}

\end{maplegroup}
\begin{maplegroup}
Selection, sets, lists

\begin{mapleinput}
\mapleinline{active}{1d}{op( int(f(x),x) ); op( [1,2,3] );}{%
}
\end{mapleinput}

\mapleresult
\begin{maplelatex}
\mapleinline{inert}{2d}{f(x), x;}{%
\[
\mathrm{f}(x), \,x
\]
}
\end{maplelatex}

\begin{maplelatex}
\mapleinline{inert}{2d}{1, 2, 3;}{%
\[
1, \,2, \,3
\]
}
\end{maplelatex}

\end{maplegroup}
\begin{maplegroup}
\begin{mapleinput}
\mapleinline{active}{1d}{lhs(x=1); rhs(f(x)=sin(x));}{%
}
\end{mapleinput}

\mapleresult
\begin{maplelatex}
\mapleinline{inert}{2d}{x;}{%
\[
x
\]
}
\end{maplelatex}

\begin{maplelatex}
\mapleinline{inert}{2d}{sin(x);}{%
\[
\mathrm{sin}(x)
\]
}
\end{maplelatex}

\end{maplegroup}
\begin{maplegroup}
\begin{mapleinput}
\mapleinline{active}{1d}{\{2,1\}; [2,1];}{%
}
\end{mapleinput}

\mapleresult
\begin{maplelatex}
\mapleinline{inert}{2d}{\{1, 2\};}{%
\[
\{1, \,2\}
\]
}
\end{maplelatex}

\begin{maplelatex}
\mapleinline{inert}{2d}{[2, 1];}{%
\[
[2, \,1]
\]
}
\end{maplelatex}

\end{maplegroup}
\begin{maplegroup}
\begin{mapleinput}
\mapleinline{active}{1d}{seq( x[i], i=1..5 ); x[i] $ i=1..5;}{%
}
\end{mapleinput}

\mapleresult
\begin{maplelatex}
\mapleinline{inert}{2d}{x[1], x[2], x[3], x[4], x[5];}{%
\[
{x_{1}}, \,{x_{2}}, \,{x_{3}}, \,{x_{4}}, \,{x_{5}}
\]
}
\end{maplelatex}

\begin{maplelatex}
\mapleinline{inert}{2d}{x[1], x[2], x[3], x[4], x[5];}{%
\[
{x_{1}}, \,{x_{2}}, \,{x_{3}}, \,{x_{4}}, \,{x_{5}}
\]
}
\end{maplelatex}

\end{maplegroup}
\begin{maplegroup}
\begin{mapleinput}
\mapleinline{active}{1d}{\{1,2\} union \{3,4\}; \{1,2\} intersect \{1,3\};}{%
}
\end{mapleinput}

\mapleresult
\begin{maplelatex}
\mapleinline{inert}{2d}{\{1, 2, 3, 4\};}{%
\[
\{1, \,2, \,3, \,4\}
\]
}
\end{maplelatex}

\begin{maplelatex}
\mapleinline{inert}{2d}{\{1\};}{%
\[
\{1\}
\]
}
\end{maplelatex}

\end{maplegroup}
\begin{maplegroup}
\begin{mapleinput}
\mapleinline{active}{1d}{f := x -> x**2:
map(f,[x,y]); map(f,x+(y*z)); map(f,x*(y+z));}{%
}
\end{mapleinput}

\mapleresult
\begin{maplelatex}
\mapleinline{inert}{2d}{[x^2, y^2];}{%
\[
[x^{2}, \,y^{2}]
\]
}
\end{maplelatex}

\begin{maplelatex}
\mapleinline{inert}{2d}{x^2+y^2*z^2;}{%
\[
x^{2} + y^{2}\,z^{2}
\]
}
\end{maplelatex}

\begin{maplelatex}
\mapleinline{inert}{2d}{x^2*(y+z)^2;}{%
\[
x^{2}\,(y + z)^{2}
\]
}
\end{maplelatex}

\end{maplegroup}
\begin{maplegroup}
Logical operators

\begin{mapleinput}
\mapleinline{active}{1d}{(x>3) and (y<=1);}{%
}
\end{mapleinput}

\mapleresult
\begin{maplelatex}
\mapleinline{inert}{2d}{3-x < 0 and y-1 <= 0;}{%
\[
3 - x < 0\ \textbf{and}\ y - 1\leq 0
\]
}
\end{maplelatex}

\end{maplegroup}
\begin{maplegroup}
\begin{mapleinput}
\mapleinline{active}{1d}{subs(\{x=5,y=0\},\%);}{%
}
\end{mapleinput}

\mapleresult
\begin{maplelatex}
\mapleinline{inert}{2d}{-2 < 0 and -1 <= 0;}{%
\[
-2 < 0\ \textbf{and}\ -1\leq 0
\]
}
\end{maplelatex}

\end{maplegroup}
\begin{maplegroup}
\begin{mapleinput}
\mapleinline{active}{1d}{eval(\%);}{%
}
\end{mapleinput}

\mapleresult
\begin{maplelatex}
\mapleinline{inert}{2d}{true;}{%
\[
\mathit{true}
\]
}
\end{maplelatex}

\end{maplegroup}
\begin{maplegroup}
Inert operators

\begin{mapleinput}
\mapleinline{active}{1d}{Sum(1/2^i,i=1..k) = sum(1/2^i,i=1..k);}{%
}
\end{mapleinput}

\mapleresult
\begin{maplelatex}
\mapleinline{inert}{2d}{Sum(1/(2^i),i = 1 .. k) = -2*(1/2)^(k+1)+1;}{%
\[
{\displaystyle \sum _{i=1}^{k}} \,{\displaystyle \frac {1}{2^{i}}
} = - 2\,({\displaystyle \frac {1}{2}} )^{(k + 1)} + 1
\]
}
\end{maplelatex}

\end{maplegroup}
\begin{maplegroup}
\begin{mapleinput}
\mapleinline{active}{1d}{Limit(Sum(1/2^i,i=1..k),k=infinity) =
limit(sum(1/2^i,i=1..k),k=infinity);}{%
}
\end{mapleinput}

\mapleresult
\begin{maplelatex}
\mapleinline{inert}{2d}{Limit(Sum(1/(2^i),i = 1 .. k),k = infinity) = 1;}{%
\[
{\displaystyle \lim _{k\rightarrow \infty }} \,{\displaystyle 
\sum _{i=1}^{k}} \,{\displaystyle \frac {1}{2^{i}}} =1
\]
}
\end{maplelatex}

\end{maplegroup}
\begin{maplegroup}
\begin{mapleinput}
\mapleinline{active}{1d}{value(Limit(Sum(1/2^i,i=1..k),k=infinity));}{%
}
\end{mapleinput}

\mapleresult
\begin{maplelatex}
\mapleinline{inert}{2d}{1;}{%
\[
1
\]
}
\end{maplelatex}

\end{maplegroup}
\begin{maplegroup}
Float evaluation

\begin{mapleinput}
\mapleinline{active}{1d}{evalf(Pi,70);}{%
}
\end{mapleinput}

\mapleresult
\begin{maplelatex}
\mapleinline{inert}{2d}{3.1415926535897932384626433832795028841971693993751058209749445923078
16;}{%
\[
3.141592653589793238462643383279502884197169399375105820974944592307816
\]
}
\end{maplelatex}

\end{maplegroup}
\begin{maplegroup}
Matrices

\begin{mapleinput}
\mapleinline{active}{1d}{with(linalg): A:=matrix([[1,2,4],[2,1,2],[4,2,1]]);}{%
}
\end{mapleinput}

\mapleresult
\begin{maplettyout}
Warning, new definition for norm
\end{maplettyout}

\begin{maplettyout}
Warning, new definition for trace
\end{maplettyout}

\begin{maplelatex}
\mapleinline{inert}{2d}{A := matrix([[1, 2, 4], [2, 1, 2], [4, 2, 1]]);}{%
\[
A :=  \left[ 
{\begin{array}{rrr}
1 & 2 & 4 \\
2 & 1 & 2 \\
4 & 2 & 1
\end{array}}
 \right] 
\]
}
\end{maplelatex}

\end{maplegroup}
\begin{maplegroup}
\begin{mapleinput}
\mapleinline{active}{1d}{det(A);}{%
}
\end{mapleinput}

\mapleresult
\begin{maplelatex}
\mapleinline{inert}{2d}{9;}{%
\[
9
\]
}
\end{maplelatex}

\end{maplegroup}
\begin{maplegroup}
\begin{mapleinput}
\mapleinline{active}{1d}{inverse(A);}{%
}
\end{mapleinput}

\mapleresult
\begin{maplelatex}
\mapleinline{inert}{2d}{matrix([[-1/3, 2/3, 0], [2/3, -5/3, 2/3], [0, 2/3, -1/3]]);}{%
\[
 \left[ 
{\begin{array}{ccc}
{\displaystyle \frac {-1}{3}}  & {\displaystyle \frac {2}{3}}  & 
0 \\ [2ex]
{\displaystyle \frac {2}{3}}  & {\displaystyle \frac {-5}{3}}  & 
{\displaystyle \frac {2}{3}}  \\ [2ex]
0 & {\displaystyle \frac {2}{3}}  & {\displaystyle \frac {-1}{3}
} 
\end{array}}
 \right] 
\]
}
\end{maplelatex}

\end{maplegroup}
\normalsize\verb+$+


\chapter{Maple II}

\section{Examples}

\subsection{Generating a Julia set}

\verbon\begin{verbatim}
# file "julia.map"

restart:

julia:=proc(c,s,file)
  local x,y,z,i,l:
  l:=[]:z:=0:
  for x from -3/2*s to 3/2*s do print(x): for y from -3/2*s to 3/2*s do
    z:=x/s+I*y/s:
    for i from 0 while i<100 and abs(Re(z))<2 and abs(Im(z))<2 do z:=z**2+c: od:
    if i=100 then l:=[op(l),[x,y]]: fi:
  od:od:

  save l,file:    
end:

time(julia(0.3-0.5*I,200,"l6.map"));

with(plots):
read "l6.map":
plotsetup(ps,plotoutput="julia3.ps",plotoptions="noborder"):
listplot(l,style=POINT,axes=NONE,symbol=POINT,scaling=CONSTRAINED);
plotsetup(default):
\end{verbatim}
\verboff

\begin{figure}[t]\center
\includegraphics[scale=0.7,angle=25]{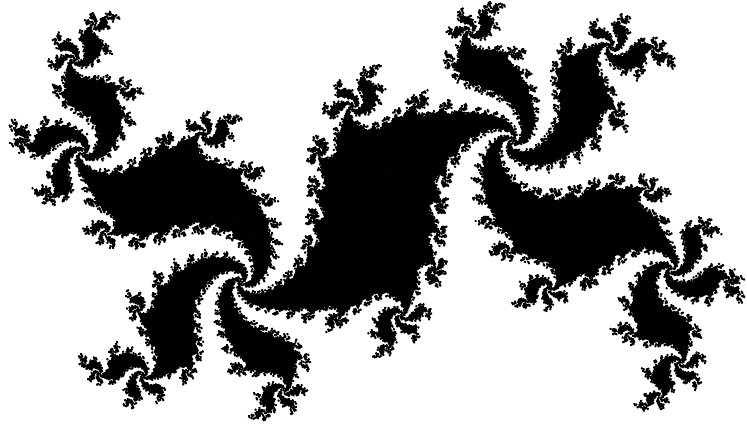}
\caption{The Julia set for $z \to z^2 + (0.3\!-\!0.5i)$. (File \emph{julia3.ps}).}
\end{figure}

\section{Advanced structures}

\subsection{Loops and conditions}

The syntax for loop and conditional commands are

\tton
if {\it boolean} then {\it statements} [elif {\it boolean} then {\it statements}] [else {\it statements}] fi;\\
for {\it name} from {\it start} [by {\it step}] to {\it stop} do {\it statements} od;\\
for {\it name} from {\it start} [by {\it step}] while {\it boolean} do {\it statements} od;\\
for {\it name} in {\it list} do {\it statements} od;\\
while {\it boolean} do {\it statements} od;\\
\ttoff

\subsection{Procedures}\label{Mapleproc}

The syntax for defining procedures in Maple is

\tton
{\it name} := proc ({\it parameters}[::{\it types}]) local {\it names}; global {\it names}; {\it statements} end;
\ttoff

The last of the {\it statements} needs no terminator and its value is
automatically the return value of the procedure. One can specify the
type of each parameter by adding {\tt ::{\it type}} to the parameters
name. Example:

\tton
f := proc(n::integer) factorial(n) end;
\ttoff

only excepts integers as parameters. Error messages or warnings can be
produced with {\tt ERROR} and {\tt WARNING}. To use other procedures
within a procedure it is sometimes useful to \emph{catch} their error
messages to prevent the procedure to be aborted. This is done be using
the {\tt traperror} command.

\subsection{Types}\label{Mapletypes}

There are innumerous different types defined in Maple. The main point
of these is that procedures and operators (also user defined ones) can
check if the given parameters have the correct type and throw error
messages if not. E.g., if you pass an integer to the display command
{\tt display(5);} it says:

\tton
Error, (in display) invalid argument, 5, must be a plot structure, or a list/set/array thereof
\ttoff

This error message if quite useful. Also, if you want to tinker plot
structures yourself and pass them to display, it is of course useful
to find the exact description/syntax of the {\tt plot} type in Maple's
help page {\tt plot, structure}. With this knowledge you could also
manipulate plots. 

The commands

\tton
whattype({\it expr});\\
type({\it expr},{\it type});\\
convert({\it expr},{\it type});
\ttoff

query the type of {\it expr}, check if {\it expr} has the type {\it
  type}, and convert {\it expr} such as to have the type {\it type},
respectively. See the help for the {\tt type} for a list of all types
Maple provides.

\section{Export / import}

For export and import there are four groups of commands. The first is
very useful to store expressions. They are:

\tton
save {\it names},"{\it filename}";\\
read "{\it filename}";
\ttoff

Example:

\tton
f:=diff(sin(x),x);\\
save f,"sample.mai";\\
restart;\\
read "sample.mai";
\ttoff

Here, the {\tt save} command produces the file {\tt sample.mai} which
simply reads:

\tton
f := cos(x);
\ttoff

The following {\tt restart} clears all assignments and cleans the
memory buffer. The {\tt read} command read the file as if it were
typed in.

The second group of commands is similar to the {\tt out} command in
Reduce and only write into files. The commands are

\tton
writeto("{\it filename}");\\
appendto("{\it filename}");\\
writeto(terminal);
\ttoff

They send all Maple's output to the specified file. {\tt
  writeto(terminal);} closes the file and redirects the output to the
worksheet.

The third group of commands allows to read and write ordinary ascii
data sheets as. Lists are used as buffer of such files:

\tton
readdata("{\it filename}" [, {\it format\_list}] [, {\it size\_of\_one\_data\_tuple}]);\\
writedata("{\it filename}", {\it list});
\ttoff

{\tt readdata} returns the data as list. The {\it format\_list} can,
e.g., be {\tt [integer,float,integer]} for producing 3-tuples of data.

The fourth group of commands implements the syntax of the programming
language C. They are

\tton
f := fopen("{\it filename}",READ|WRITE|APPEND,TEXT|BINARY);\\
fprintf(f,{\it structure\_string},{\it exprs});\\
fscanf(f,{\it structure\_string},{\it variables});\\
fclose(f);
\ttoff

The meaning should be clear if one knows C. The commands {\tt feof,
  fflush, fremove, filepos} also belong to this group. With these
commands you can produce files with arbitrary content.

\subsection{LaTeX, Fortran, and C}

The commands {\tt latex({\it expr}), fortran({\it expr})}, and {\tt
  C({\it expr})} are nice facilities to export to LaTeX, Fortran of
the programming language C. See these examples:

\verbon\begin{verbatim}
> restart: readlib(C):

> latex(Limit(Sum(1/2^i,i=1..k),k=infinity));
lim _{k\rightarrow \infty }\sum _{i=1}^{k}\left ({2}^{i}\right )^{-1}


> f:= x -> Pi*ln(x^2)-sqrt(2)*ln(x^2)^2;

> C(f,optimized);
/* The options were    : operatorarrow */
#include <math.h>
double f(x)
double x;
{
  double t2;
  double t1;
  double t4;
  double t5;
  {
    t1 = x*x;
    t2 = log(t1);
    t4 = sqrt(2.0);
    t5 = t2*t2;
    return(0.3141592653589793E1*t2-t4*t5);
  }
}

> fastgrow:=proc(k)
>   local x,i;
>   x:=1;
>   for i from 1 to k do x:=x+x**x; od;
>   x;
> end;

> fortran(fastgrow);
      real function fastgrow(k)
      real k

      integer i
      real x

        x = 1
        do 1000 i = 1,k,1
          x = x+x**x
1000    continue
        fastgrow = x
        return
      end
\end{verbatim}
\verboff

A possibility to export whole Maple worksheet to Latex is with the
menu \emph{File -- Export As -- LaTeX}. I used this method in this
paper. This will produce a LaTeX document reproducing the appearance
of the worksheet. The Maple style files have to be installed for this.
One can get them at\\ 
\verb+http://www.maplesoft.com/latex_patch.html+\\ and provisionally
copy them in the same directory as the LaTeX file.

\subsection{Plotting}

Again, Maple offers innumerous commands to generate plots. Here, I
present only those that I find most elementary:

\tton
plot({\it function\_list} [,{\it horizontal\_range}] [,{\it vertical\_range}] [,{\it plot\_options}]);\\
textplot(\{[{\it x\_coord},{\it y\_coord},"{\it text}"],..\} [,{\it plot\_options}]);\\
display({\it plot\_structure\_list} [,{\it plot\_options}]);\\
with(plots);\\
odeplot({\it numeric\_dsolve\_output},{\it function\_list},{\it x\_range} [,{\it plot\_options}]);\\
plotsetup({\it device\_type} [,{\it options}]);
\ttoff

For handling plots it is important to realize that all plot commands
return a plot structure of type {\tt PLOT}. See this example:

\verbon\begin{verbatim}
plt:=plot(sin(x),x=-Pi..Pi):
save plt,"sample.plt":
restart:
read "sample.plt":
display(plt);
\end{verbatim}
\verboff

\begin{figure}[t]\center
\includegraphics[scale=0.4,angle=-90]{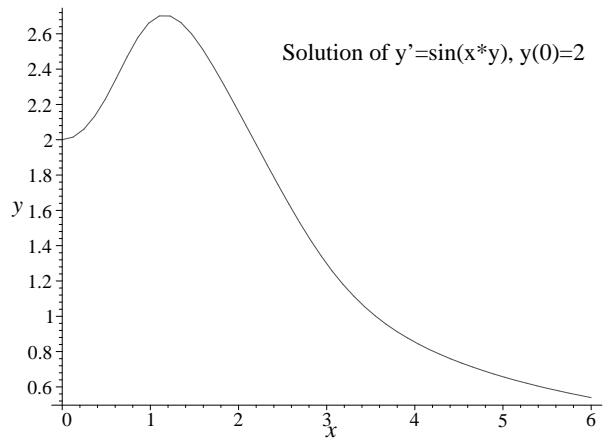}
\caption{Plot produced by Maple}
\label{Fmapleplot}
\end{figure}

Here, we assign the plot of $\sin(x)$ to the variable {\tt plt} and
store this variable in the {\tt sample.plt} file. This is very useful
when it takes long to produce plots (e.g.\ with {\tt odeplot}!). The
following commands produce the postscript in figure \ref{Fmapleplot}
and write it in the file {\tt mapplot.ps}.

\verbon\begin{verbatim}
with(plots):
solution := dsolve({ diff(y(x),x) = sin(x*y(x)),y(0)=2},y(x),type=numeric):
plt := odeplot(solution,[x,y(x)],0..6,labels=[x,y],labelfont=[TIMES,ITALIC,22]):
txt := textplot([2.5,2.5,"Solution of y'=sin(x*y), y(0)=2"],align=RIGHT,font=[TIMES,ROMAN,22]):
plotsetup(ps,plotoutput="mapplot.ps",plotoptions="noborder"):
display({plt,txt},axesfont=[TIMES,ROMAN,18]);
plotsetup(default):
\end{verbatim}
\verboff

\chapter{Tensor type calculi on Reduce and Maple}

\section{Maple}

\subsection{The {\tt tensor} package}

Maple provides a the {\tt tensor} package to implement the tensor
calculus. This package is strongly oriented to perform the standard
calculations in Einsteinian gravity, i.e.\ to calculate the
Christoffel symbol, the Riemannian curvature tensor, the Ricci tensor,
etc. from a given symmetric metric. For this special purpose the
package is quite useful. Of course, the package provides also
elementary operations to work with tensors: adding tensors {\tt
  lin\_com}, the tensor product {\tt prod}, raising, lowering,
contracting, and (anti-) symmetrizing indices {\tt raise, lower,
  contract, antisymmetrize, symmetrize}. However, as one can see from
table \ref{Ttensorcoms}, the syntax of these commands is not at all in
analogy to the usual notation of tensor calculus: To setup a general
3rd rank tensor $A$ in 2D space with two covariant and one
contravariant indices one has to declare

\verbon\begin{verbatim}
restart:with(tensor):
A:=create([-1,-1,1],array(
  [[[a.1.1.1(x.1,x.2),a.1.1.2(x.1,x.2)],[a.1.2.1(x.1,x.2),a.1.2.2(x.1,x.2)]],
   [[a.2.1.1(x.1,x.2),a.2.1.2(x.1,x.2)],[a.2.2.1(x.1,x.2),a.2.2.2(x.1,x.2)]]]
));
\end{verbatim}
\verboff

To specify the tensor type, {\tt '-1'} means a covariant index whereas
{\tt '1'} means a contravariant (in contrast to what is said in the
help page of {\tt create}). Maple starts counting indices from 1. As
you see, one has to introduce general names for all components oneself
(that include the dependency on coordinates). The following examples
show other tensor manipulations:

\begin{tabular}{p{20ex}|p{61ex}}
usual notation & {\tt tensor} package syntax \spc\\
\hline
$5\, A_{ij}{}^k + B_{ij}{}^k + 2\, C_{ij}{}^k$ & {\tt lin\_com(5,A,B,2,c); } \spc\\
$A_{ijk}\, B_{lm}{}^i$ & {\tt prod(A,B,[1,3]); } \spc\\
$A_{ij}{}^i$ & {\tt contract(A,[1,3]); } \spc\\
$\del_i\, A_{jk}$ & {\tt partial\_diff(A,[x.1,x.2,x.3,x.4]); }
\end{tabular}

In contrast, we will see the Excalc allows tensor manipulations that
are a direct mirror of the usual notation. The {\tt entermetric}
command allows to enter a metric interactively. Personally, I find
this unacceptable for a CA system.

\begin{table}[t]
\begin{tabular}{p{15ex}|p{66ex}}

declarations & {\tt create ({\it tensor\_type},{\it
  components\_array});~~ entermetric } \spc\\ 

elementary ops & {\tt lin\_com ({\it
    coefficiants\_and\_tensors});\newline prod ({\it 1st\_tensor},{\it
    2nd\_tensor},{\it index\_pairs\_to\_contract});\newline raise
  lower ({\it metric},{\it tensor},{\it indices});\newline contract
  ({\it tensor},{\it index\_pairs}); \newline symmetrize
  antisymmetrize ({\it tensor},{\it index\_array}); \newline
  get\_compts get\_char get\_rank ({\it tensor});} \spc\\ 

differentiation & {\tt partial\_diff exterior\_diff ({\it tensor},{\it
    coord\_array});\newline cov\_diff ({\it tensor},{\it coord\_array},{\it
    Christoffel\_symbol});\newline Lie\_diff directional\_diff ({\it
    tensor},{\it vector},{\it coord\_array}); } \spc\\ 

more ops & {\tt compare invert dual permute\_indices
  exterior\_prod act conj frame commutator } \spc\\ 

certain tensors & {\tt Levi\_Civita d1metric d2metric } \spc\\ 

coord trans & {\tt Jacobian change\_basis transform } \spc\\ 

GR tensors & {\tt tensorsGR display\_allGR displayGR Christoffel1
  Christoffel2 Einstein Ricci Ricciscalar Riemann Weyl connexF
  RiemannF invars petrov } \spc\\ 

Newmann-P. & {\tt convertNP npcurve npspin } \spc\\ 

extract eqns & {\tt Killing\_eqns geodesic\_eqns } \spc\\
\end{tabular}
\caption{
The commands of Maple's tensor package. The syntax is only specified for elementary commands. }
\label{Ttensorcoms}
\end{table}

Finally, we show how simple it is to calculate the standard GR tensors
with the {\tt tensor} package for a given metric. In the example we
derive the Schwarzschild metric as solution of the Einstein equation
for a general static, spherically symmetric ansatz.

\footnotesize\begin{maplegroup}
\begin{mapleinput}
\mapleinline{active}{1d}{restart:with(tensor):}{%
}
\end{mapleinput}

\end{maplegroup}
\begin{maplegroup}
\begin{mapleinput}
\mapleinline{active}{1d}{gdd:=create([-1,-1],array(symmetric,[
[f(r),0,0,0],
[0,-g(r)/f(r),0,0],
[0,0,-r^2,0],
[0,0,0,-r^2*sin(theta)^2]
]));}{%
}
\end{mapleinput}

\mapleresult
\begin{maplelatex}
\mapleinline{inert}{2d}{gdd := TABLE([compts = matrix([[f(r), 0, 0, 0], [0, -g(r)/f(r), 0,
0], [0, 0, -r^2, 0], [0, 0, 0, -r^2*sin(theta)^2]]), index_char = [-1,
-1]]);}{%
\maplemultiline{
\mathit{gdd} := \mathrm{table(}[ \\
\mathit{compts}= \left[ 
{\begin{array}{cccc}
\mathrm{f}(r) & 0 & 0 & 0 \\
0 &  - {\displaystyle \frac {\mathrm{g}(r)}{\mathrm{f}(r)}}  & 0
 & 0 \\ [2ex]
0 & 0 &  - r^{2} & 0 \\
0 & 0 & 0 &  - r^{2}\,\mathrm{sin}(\theta )^{2}
\end{array}}
 \right]  \\
\mathit{index\_char}=[-1, \,-1] \\
]) }
}
\end{maplelatex}

\end{maplegroup}
\begin{maplegroup}
\begin{mapleinput}
\mapleinline{active}{1d}{tensorsGR([t,r,theta,phi],gdd,
guu,gdet,chris1,chris2,riem,ric,ric_scalar,einstein,weyl):}{%
}
\end{mapleinput}

\end{maplegroup}
\begin{maplegroup}
\begin{mapleinput}
\mapleinline{active}{1d}{eqs:=get_compts(ric);}{%
}
\end{mapleinput}

\mapleresult
\begin{maplelatex}
\mapleinline{inert}{2d}{eqs :=
matrix([[1/4*f(r)*(-2*r*g(r)*diff(f(r),`$`(r,2))+r*diff(g(r),r)*diff(f
(r),r)-4*g(r)*diff(f(r),r))/(g(r)^2*r), 0, 0, 0], [0,
-1/4*(-2*r*g(r)*diff(f(r),`$`(r,2))+r*diff(g(r),r)*diff(f(r),r)+4*diff
(g(r),r)*f(r)-4*g(r)*diff(f(r),r))/(r*f(r)*g(r)), 0, 0], [0, 0,
1/2*(2*diff(f(r),r)*r*g(r)-r*diff(g(r),r)*f(r)-2*g(r)^2+2*f(r)*g(r))/(
g(r)^2), 0], [0, 0, 0,
1/2*sin(theta)^2*(2*diff(f(r),r)*r*g(r)-r*diff(g(r),r)*f(r)-2*g(r)^2+2
*f(r)*g(r))/(g(r)^2)]]);}{%
\[
\mathit{eqs} :=  \left[ 
{\begin{array}{c}
{\displaystyle \frac {1}{4}} \,{\displaystyle \frac {\mathrm{f}(r
)\,( - 2\,r\,\mathrm{g}(r)\,({\frac {\partial ^{2}}{\partial r^{2
}}}\,\mathrm{f}(r)) + r\,({\frac {\partial }{\partial r}}\,
\mathrm{g}(r))\,({\frac {\partial }{\partial r}}\,\mathrm{f}(r))
 - 4\,\mathrm{g}(r)\,({\frac {\partial }{\partial r}}\,\mathrm{f}
(r)))}{\mathrm{g}(r)^{2}\,r}} \,, \,0\,, \,0\,, \,0 \\ [2ex]
0\,, \, - {\displaystyle \frac {1}{4}} \,{\displaystyle \frac {
 - 2\,r\,\mathrm{g}(r)\,({\frac {\partial ^{2}}{\partial r^{2}}}
\,\mathrm{f}(r)) + r\,({\frac {\partial }{\partial r}}\,\mathrm{g
}(r))\,({\frac {\partial }{\partial r}}\,\mathrm{f}(r)) + 4\,({\frac {\partial }{\partial r}}\,\mathrm{g}(r))\,\mathrm{f}(r) - 4\,\mathrm{g}(r)\,({\frac {\partial }{
\partial r}}\,\mathrm{f}(r))}{r\,\mathrm{f}(r)\,\mathrm{g}(r)}} \,, \,0\,, \,0 \\ [2ex]
0\,, \,0\,, \,{\displaystyle \frac {1}{2}} \,{\displaystyle 
\frac {2\,({\frac {\partial }{\partial r}}\,\mathrm{f}(r))\,r\,\mathrm{g}(r) - r\,({\frac {\partial }{\partial r}}\,
\mathrm{g}(r))\,\mathrm{f}(r) - 2\,\mathrm{g}(r)^{2} + 2\,\mathrm{f}(r)\,
\mathrm{g}(r)}{\mathrm{g}(r)^{2}}} \,, \,0 \\ [2ex]
0\,, \,0\,, \,0\,, \,{\displaystyle \frac {1}{2}} \,
{\displaystyle \frac {\mathrm{sin}(\theta )^{2}\,(2\,({\frac {
\partial }{\partial r}}\,\mathrm{f}(r))\,r\,\mathrm{g}(r) - r\,(
{\frac {\partial }{\partial r}}\,\mathrm{g}(r))\,\mathrm{f}(r) - 
2\,\mathrm{g}(r)^{2} + 2\,\mathrm{f}(r)\,\mathrm{g}(r))}{\mathrm{
g}(r)^{2}}} 
\end{array}}
 \right] 
\]
}
\end{maplelatex}

\end{maplegroup}
\begin{maplegroup}
\begin{mapleinput}
\mapleinline{active}{1d}{dsolve(\{eqs[1,1],eqs[2,2]\},\{f(r),g(r)\});}{%
}
\end{mapleinput}

\mapleresult
\begin{maplelatex}
\mapleinline{inert}{2d}{\{f(r) = _C2+_C3/r\}, \{g(r) = _C1\};}{%
\[
\{\mathrm{f}(r)=\mathit{\_C2} + {\displaystyle \frac {\mathit{
\_C3}}{r}} \}, \,\{\mathrm{g}(r)=\mathit{\_C1}\}
\]
}
\end{maplelatex}

\end{maplegroup}
\normalsize

\subsection{The GRTensor package}

\newpage
\footnotesize\begin{maplegroup}
\begin{mapleinput}
\mapleinline{active}{1d}{restart:
read "/home/mt/usr/maple/lib5/grii.m":
grtensor():}{%
}
\end{mapleinput}

\end{maplegroup}
\begin{maplegroup}
\begin{mapleinput}
\mapleinline{active}{1d}{makeg(ss):}{%
}
\end{mapleinput}

\emptyline
\mapleresult
\begin{verbatim}
Makeg 2.0: GRTensor metric/basis entry utility

To quit makeg, type 'exit' at any prompt.{\large }

Do you wish to enter a 1) metric [g(dn,dn)],
                       2) line element [ds],
                       3) non-holonomic basis [e(1)...e(n)], or
                       4) NP tetrad [l,n,m,mbar]?
\end{verbatim}

\emptyline
\end{maplegroup}
\begin{maplegroup}
\begin{mapleinput}
\mapleinline{active}{1d}{2:}{%
}
\end{mapleinput}

\mapleresult
\begin{verbatim}
Enter coordinates as a LIST (e.g. [r,theta,phi,t]):
\end{verbatim}

\end{maplegroup}
\begin{maplegroup}
\begin{mapleinput}
\mapleinline{active}{1d}{[t,r,theta,phi]:}{%
}
\end{mapleinput}

\mapleresult
\begin{verbatim}
Enter the line element using d[coord] to indicate differentials.
(for example,  r^2*(d[theta]^2 + sin(theta)^2*d[phi]^2)
[Type 'exit' to quit makeg]
 ds^2 = 
\end{verbatim}

\end{maplegroup}
\begin{maplegroup}
\begin{mapleinput}
\mapleinline{active}{1d}{f(r)*d[t]^2 - g(r)/f(r)*d[r]^2 - r^2*(d[theta]^2 + sin(theta)^2 *
d[phi]^2):}{%
}
\end{mapleinput}

\emptyline
\mapleresult
\begin{verbatim}
If there are any complex valued coordinates, constants or functions
for this spacetime, please enter them as a SET ( eg. { z, psi } ).
\end{verbatim}

\emptyline
\begin{verbatim}
Complex quantities [default={}]: 
\end{verbatim}

\end{maplegroup}
\begin{maplegroup}
\begin{mapleinput}
\mapleinline{active}{1d}{\{\}:}{%
}
\end{mapleinput}

\mapleresult
\begin{maplelatex}
\mapleinline{inert}{2d}{`The values you have entered are:`;}{%
\[
\mathit{The\ values\ you\ have\ entered\ are:}
\]
}
\end{maplelatex}

\begin{maplelatex}
\mapleinline{inert}{2d}{Coordinates = [t, r, theta, phi];}{%
\[
\mathit{Coordinates}=[t, \,r, \,\theta , \,\phi ]
\]
}
\end{maplelatex}

\begin{maplelatex}
\mapleinline{inert}{2d}{`Metric:`;}{%
\[
\mathit{Metric:}
\]
}
\end{maplelatex}

\begin{maplelatex}
\mapleinline{inert}{2d}{g[a]*``[b] = matrix([[f(r), 0, 0, 0], [0, -g(r)/f(r), 0, 0], [0, 0,
-r^2, 0], [0, 0, 0, -r^2*sin(theta)^2]]);}{%
\[
{g_{a}}\,{\ _{b}}= \left[ 
{\begin{array}{cccc}
\mathrm{f}(r) & 0 & 0 & 0 \\
0 &  - {\displaystyle \frac {\mathrm{g}(r)}{\mathrm{f}(r)}}  & 0
 & 0 \\ [2ex]
0 & 0 &  - r^{2} & 0 \\
0 & 0 & 0 &  - r^{2}\,\mathrm{sin}(\theta )^{2}
\end{array}}
 \right] 
\]
}
\end{maplelatex}

\begin{verbatim}
You may choose to 0) Use the metric WITHOUT saving it,
                  1) Save the metric as it is,
                  2) Correct an element of the metric,
                  3) Re-enter the metric,
                  4) Add/change constraint equations, 
                  5) Add a text description, or
                  6) Abandon this metric and return to Maple.
\end{verbatim}

\emptyline
\end{maplegroup}
\begin{maplegroup}
\begin{mapleinput}
\mapleinline{active}{1d}{0:}{%
}
\end{mapleinput}

\mapleresult
\begin{verbatim}
Calculated ds for ss (.020000 sec.) 
\end{verbatim}

\begin{maplelatex}
\mapleinline{inert}{2d}{`Default spacetime` = ss;}{%
\[
\mathit{Default\ spacetime}=\mathit{ss}
\]
}
\end{maplelatex}

\begin{maplelatex}
\mapleinline{inert}{2d}{`For the ss spacetime:`;}{%
\[
\mathit{For\ the\ ss\ spacetime:}
\]
}
\end{maplelatex}

\begin{maplelatex}
\mapleinline{inert}{2d}{Coordinates;}{%
\[
\mathit{Coordinates}
\]
}
\end{maplelatex}

\begin{maplelatex}
\mapleinline{inert}{2d}{x(up);}{%
\[
\mathrm{x}(\mathit{up})
\]
}
\end{maplelatex}

\begin{maplelatex}
\mapleinline{inert}{2d}{`x `^a = vector([t, r, theta, phi]);}{%
\[
\mathit{x\ }^{a}=[t, \,r, \,\theta , \,\phi ]
\]
}
\end{maplelatex}

\begin{maplelatex}
\mapleinline{inert}{2d}{`Line element`;}{%
\[
\mathit{Line\ element}
\]
}
\end{maplelatex}

\begin{maplelatex}
\mapleinline{inert}{2d}{` ds`^2 = f(r)*` d`*t^`2 `-g(r)*` d`*r^`2 `/f(r)-r^2*` d`*theta^`2
`-r^2*sin(theta)^2*` d`*phi^`2 `;}{%
\[
\mathit{\ ds}^{2}=\mathrm{f}(r)\,\mathit{\ d}\,t^{\mathit{2\ }}
 - {\displaystyle \frac {\mathrm{g}(r)\,\mathit{\ d}\,r^{\mathit{
2\ }}}{\mathrm{f}(r)}}  - r^{2}\,\mathit{\ d}\,\theta ^{\mathit{2
\ }} - r^{2}\,\mathrm{sin}(\theta )^{2}\,\mathit{\ d}\,\phi ^{
\mathit{2\ }}
\]
}
\end{maplelatex}

\begin{verbatim}
makeg() completed.
\end{verbatim}

\end{maplegroup}
\begin{maplegroup}
\begin{mapleinput}
\mapleinline{active}{1d}{grcalc( R(dn,dn) ):}{%
}
\end{mapleinput}

\mapleresult
\begin{verbatim}
Calculated detg for ss (.010000 sec.) 
Calculated g(up,up) for ss (.020000 sec.) 
Calculated g(dn,dn,pdn) for ss (.020000 sec.) 
Calculated Chr(dn,dn,dn) for ss (.020000 sec.) 
Calculated Chr(dn,dn,up) for ss (.040000 sec.) 
Calculated R(dn,dn) for ss (.040000 sec.) 
\end{verbatim}

\begin{maplelatex}
\mapleinline{inert}{2d}{`CPU Time ` = .150;}{%
\[
\mathit{CPU\ Time\ }=.150
\]
}
\end{maplelatex}

\end{maplegroup}
\begin{maplegroup}
\begin{mapleinput}
\mapleinline{active}{1d}{grdisplay( R(dn,dn) ):}{%
}
\end{mapleinput}

\mapleresult
\begin{maplelatex}
\mapleinline{inert}{2d}{`For the ss spacetime:`;}{%
\[
\mathit{For\ the\ ss\ spacetime:}
\]
}
\end{maplelatex}

\begin{maplelatex}
\mapleinline{inert}{2d}{`Covariant Ricci`;}{%
\[
\mathit{Covariant\ Ricci}
\]
}
\end{maplelatex}

\begin{maplelatex}
\mapleinline{inert}{2d}{R(dn,dn);}{%
\[
\mathrm{R}(\mathit{dn}, \,\mathit{dn})
\]
}
\end{maplelatex}

\begin{maplelatex}
\mapleinline{inert}{2d}{`R `[a]*``[b] =
matrix([[-1/4*f(r)*(diff(f(r),r)*diff(g(r),r)*r-2*diff(f(r),`$`(r,2))*
g(r)*r-4*g(r)*diff(f(r),r))/(g(r)^2*r), 0, 0, 0], [0,
1/4*(-2*diff(f(r),`$`(r,2))*g(r)*r+diff(f(r),r)*diff(g(r),r)*r+4*diff(
g(r),r)*f(r)-4*g(r)*diff(f(r),r))/(f(r)*g(r)*r), 0, 0], [0, 0,
1/2*(-2*diff(f(r),r)*r*g(r)+f(r)*r*diff(g(r),r)+2*g(r)^2-2*f(r)*g(r))/
(g(r)^2), 0], [0, 0, 0,
1/2*sin(theta)^2*(-2*diff(f(r),r)*r*g(r)+f(r)*r*diff(g(r),r)+2*g(r)^2-
2*f(r)*g(r))/(g(r)^2)]]);}{%
\[
{\mathit{R\ }_{a}}\,{\ _{b}}= \left[ 
{\begin{array}{c}
 - {\displaystyle \frac {1}{4}} \,{\displaystyle \frac {\mathrm{f
}(r)\,(({\frac {\partial }{\partial r}}\,\mathrm{f}(r))\,(
{\frac {\partial }{\partial r}}\,\mathrm{g}(r))\,r - 2\,({\frac {
\partial ^{2}}{\partial r^{2}}}\,\mathrm{f}(r))\,\mathrm{g}(r)\,r
 - 4\,\mathrm{g}(r)\,({\frac {\partial }{\partial r}}\,\mathrm{f}
(r)))}{\mathrm{g}(r)^{2}\,r}} \,, \,0\,, \,0\,, \,0 \\ [2ex]
0\,, \,{\displaystyle \frac {1}{4}} \,{\displaystyle \frac { - 2
\,({\frac {\partial ^{2}}{\partial r^{2}}}\,\mathrm{f}(r))\,
\mathrm{g}(r)\,r + ({\frac {\partial }{\partial r}}\,\mathrm{f}(r))\,({\frac {
\partial }{\partial r}}\,\mathrm{g}(r))\,r + 4\,({\frac {\partial }{\partial r}}\,\mathrm{g}(r))\,\mathrm{
f}(r) - 4\,\mathrm{g}(r)\,({\frac {\partial }{\partial r}}\,
\mathrm{f}(r))}{\mathrm{f}(r)\,\mathrm{g}(r)\,r}} \,, \,0\,, \,0 \\ [2ex]
0\,, \,0\,, \,{\displaystyle \frac {1}{2}} \,{\displaystyle 
\frac { - 2\,({\frac {\partial }{\partial r}}\,\mathrm{f}(r))\,r\,\mathrm{g}(r) + \mathrm{f}(r)\,r\,({\frac {\partial }{\partial r}}\,\mathrm{g}(r)) + 2\,\mathrm{
g}(r)^{2} - 2\,\mathrm{f}(r)\,\mathrm{g}(r)}{\mathrm{g}(r)^{2}}} 
\,, \,0 \\ [2ex]
0\,, \,0\,, \,0\,, \,{\displaystyle \frac {1}{2}} \,
{\displaystyle \frac {\mathrm{sin}(\theta )^{2}\,( - 2\,({\frac {
\partial }{\partial r}}\,\mathrm{f}(r))\,r\,\mathrm{g}(r) + \mathrm{f}(r)\,r\,({\frac {\partial }{\partial r}}\,\mathrm{g}(r)) + 2\,\mathrm{
g}(r)^{2} - 2\,\mathrm{f}(r)\,\mathrm{g}(r))}{\mathrm{g}(r)^{2}}
} 
\end{array}}
 \right] 
\]
}
\end{maplelatex}

\end{maplegroup}
\begin{maplegroup}
\begin{mapleinput}
\mapleinline{active}{1d}{eqs:=\{grcomponent(R (dn,dn), [1,1]),grcomponent(R (dn,dn),
[2,2])\}:}{%
}
\end{mapleinput}

\end{maplegroup}
\begin{maplegroup}
\begin{mapleinput}
\mapleinline{active}{1d}{dsolve(eqs,\{f(r),g(r)\});}{%
}
\end{mapleinput}

\mapleresult
\begin{maplelatex}
\mapleinline{inert}{2d}{\{f(r) = _C2+_C3/r\}, \{g(r) = _C1\};}{%
\[
\{\mathrm{f}(r)=\mathit{\_C2} + {\displaystyle \frac {\mathit{
\_C3}}{r}} \}, \,\{\mathrm{g}(r)=\mathit{\_C1}\}
\]
}
\end{maplelatex}

\end{maplegroup}
\begin{maplegroup}
\begin{mapleinput}
\end{mapleinput}

\end{maplegroup}
\normalsize

\verbon\begin{verbatim}
restart:
read "/files/home/part3/mt/usr/maple/lib5/grii.m":
grtensor():
makeg(ss):
2:
[t,r,theta,phi]:
f(r)*d[t]^2 - g(r)/f(r)*d[r]^2 - r^2*(d[theta]^2 + sin(theta)^2 * d[phi]^2):
{}:
0:
grcalc( R(dn,dn) ):
grdisplay( R(dn,dn) ):
eqs:={grcomponent(R (dn,dn), [1,1]),grcomponent(R (dn,dn), [2,2])}:
dsolve(eqs,{f(r),g(r)}):
latex({%});
\end{verbatim}
\verboff

\mapon
\left \{\left \{g(r)={\it \_C1}\right \},\left \{f(r)={\it \_C2}+{\frac {{\it \_C3}}{r}}\right \}\right \}
\mapoff

\section{Excalc}

Who doubts about the beauty of the exterior calculus?! Reduce offers a
package Excalc that provides very handy methods to handle exterior
forms as well as tensors. Fortunately, this package implements
elementary declarations and operators that let all freedom to the user
to define quantities - it does not offer complete GR routines as the
{\tt tensor} of {\tt GRtensor} packages in Maple that impose their
notations and conventions. Excalc \emph{really} generally implements
the tensor \emph{and} exterior calculus. Tensors can also be handled
by Reduce without the Excalc package -- you can find some explanation
for this in the appendix. However, since with Excalc it is a lot
easier and more beautiful, I advice to use the Excalc package to
manipulate tensors.

The Excalc package is loaded with {\tt load\_package excalc\$}. It
provides the commands and operators summarized in table
\ref{Texcalccoms}. The table should explain most commands. There is
online help available at \cite{excalcpage}. We now give more detailed
explanations on the basic commands.

\begin{table}[t]\center
\begin{tabular}{p{25ex}|p{61ex}}
declare forms/tensors & \tt pform {\it name}[({\it tensor\_indices})]={\it degree};\\
declare vectors & \tt tvector {\it names};\\
declare dependence & \tt fdomain {\it name}={\it name}({\it dependency});\\
tensor symmetries & \tt index\_symmetries {\it tensor}:\ [symmetric [in {\it index\_pairs}]] [antisymmetric [in {\it index\_pairs}]];\\
get form degree & \tt exdegree({\it form});\\
\hline
exterior derivative & \tt d ({\it form});\\
partial differentiation & \tt @ ({\it form},{\it coord});\\
inner product & \tt {\it vector} \_| {\it form};\\
hodge dual & \tt \# ({\it form});\\
Lie derivative & \tt {\it vector} |\_ {\it form};\\
variational derivative & \tt vardf ({\it form},{\it non\_tensor\_form});\\
\hline
declare coframe & \tt coframe {\it coframe\_def} [with metric {\it metric\_def} | with signature {\it signature}];\\
declare frame & \tt frame {\it name};\\
display coframe & \tt displayframe;\\
indices & \tt indexrange {\it numbers\_or\_names};\\
set space dimension & \tt spacedim {\it number};\\
\hline
tangent vector & \tt @ ({\it coord});\\
epsilon tensor & \tt eps ({\it indices});\\
Christoffel symbol & \tt riemannconx {\it name};\\
\hline
others & \tt noether keep nosum renosum noxpnd forder remforder xpnd 
\end{tabular}
\caption{
  The commands of Excalc. A {\it form} is in general an exterior form
  declared with {\tt pform} which can have tensor indices also. For
  the meaning of {\it index\_pairs, coframe\_def, metric\_def,} and
  {\it signature} see the text.}
\label{Texcalccoms}
\end{table}

The most important declaration is {\tt pform} which allows to
introduce exterior forms with additional tensor indices. E.g., with
the following we introduce a 2rd rank tensor {\tt t} ($t^{ij}$), a 4th
rank tensor {\tt r} ($r^{ijkl}$), and a 2-form {\tt p} with one index
($p^i \in \L^2$):

\tton
pform t(i,j)=0,r(i,j,k,l)=0,p(i)=2\$
\ttoff

After this declaration, these tensors/forms can be refered to by {\tt
  t({\it 1st\_index\_name},{\it 2nd\_index\_name})}, etc., i.e.\ we
can use arbitrary names for the indices. Specifying the index range
lets Excalc run the index names over this range in all expression.

Putting a minus sign in front of the index name lowers the index,
without a minus sign the index is a contravariant index. {\bf Excalc
  knows the summation convention} summing over identical lower and
upper tensor indices. E.g., the following examples calculate the trace
$t_i{}^i$ and perform the assignment $t_{ij}:=r_{kij}{}^k$:

\verbon\begin{verbatim}
1: load_package excalc$
*** ^ redefined 

2: pform t(i,j)=0,r(i,j,k,l)=0,p(i)=2$

3: t(-i,i);

   i
t
 i 

4: t(i,j):=r(-k,i,j,k);

 i j       i j k
t    := r
         k 

5: indexrange 0,1,y,z$

6: t(-i,i);

   0      1      y      z
t    + t    + t    + t
 0      1      y      z 

7: t(0,0):=a$ t(0,0);

a

9: t (-0,0);

   0
t
 0 
\end{verbatim}
\verboff

If we have defined a metric Excalc automatically calculates the
(contra-)covariant components from assignments to (co-)contravariant
components. For a diagonal metric, the response to input 9 would then
be $g_{00}\, a$. Specifying the metric structure is a little bit
involved. Roughly, the syntax to introduce a frame and metric is

\tton
coframe {\it coframe\_def} [with metric {\tt metric\_def} | with signature {\it signature}];
\ttoff

The following examples should make clear what the parameters mean:

\verbon\begin{verbatim}
% file "exc1"
load_package excalc$

% symbolic 3D coframe with Euclidean metric:
coframe o(0),o(1),o(2)$

% polar coframe with Euclidean metric:
coframe o(r)=d r,o(p)=r * d phi$    

% Schwarzschild metric in anholonomic/diagonalized form:
coframe o(0)=f * d t,
        o(1)=1/f * d r,
        o(2)=r * d theta,
        o(3)=r*sin(theta) * d phi
with signature (+1,-1,-1,-1)$          

% symbolic 4D coframe with arbitrary metric:
operator gll,o$         % introduce general notations
for i:=0:3 do for j:=i+1:3 do gll(i,j):=gll(j,i)$     % symmetrize metric
let { metric_def => for i:=0:3 sum for j:=0:3 sum gll(i,j)*o(i)*o(j) };
coframe o(0),o(1),o(2),o(3) with metric g = metric_def$
                                    
% holonomic 4D coframe with arbitrary metric and coordinates x(a):
operator gll,o$
for i:=0:3 do for j:=i+1:3 do gll(i,j):=gll(j,i)$
let { metric_def => for i:=0:3 sum for j:=0:3 sum gll(i,j)*o(i)*o(j) };
coframe o(0)=d x0,o(1)=d x1,o(2)=d x2,o(3)=d x3 with metric g = metric_def$
pform x(i)=0$           % introduce coordinates as indexed 0-form defined as:
x(-0):=x0$ x(-1):=x1$ x(-2):=x2$ x(-3):=x3$
frame e$                % generate a frame e(-a) dual to o(a)
e(-a) _| o(-b);         %-> gll(a,b)
\end{verbatim}
\verboff

The first example introduces a coframe index by indices running over
$0,1,2$ in the Euclidean (because no metric is specified) 3D space.
The second example introduces alphabetic(!) indices running over $r,p$
and defines the anhonomic coframe with respect to holonomic coordinate
derivatives ($dr$,$d\phi$). Thus, also {\tt r} and {\tt phi} are
introduces as coordinates (e.g.\ enabling to write {\tt @ r} for the
radial vector). The 3rd example introduces the Schwarzschild geometry
with diagonal metric and anholonomic coframe.

The 4th example introduces a general coframe with arbitrary metric.
The metric (low-low) components have been predefined as symmetric
operator. Since the coframe command does not accept {\tt for}
constructions, the variable {\tt metric\_def} was used as \emph{alias}
for the metric expression. It is better to use {\tt let} rules to
predefine expressions that will be used in the coframe definition. The
5th example introduces a \emph{holonomic} coframe with arbitrary
metric, coordinate basis {\tt x(i)}, and holonomic frame {\tt @ x(a)}
which coincides with {\tt e(-a)}.

The command

\tton
frame e\$
\ttoff

introduces the name {\tt e} for the respective frame (defined via the
metric by $e_i\, \inn o^j = \d_i^j$ or $e_i\, \inn o_j = g_{ij}$). Thus,
the last example of the coframe definitions will yield

\verbon\begin{verbatim}
e(-a) _| o(-b)                \% throws  gll(a,b)
\end{verbatim}
\verboff

~

Specifying the symmetries of tensors (or tensor indices on forms)
saves Reduce from calculating redundant components and thus saves
time. Excalc provides the {\tt index\_symmetries} command for this.
Note that if no metric is specified before the {\tt index\_symmetries}
command assumes an Euclidean metric! The syntax given in the table is

\tton
index\_symmetries {\it tensor}:\ [symmetric [in {\it index\_pairs}]] [antisymmetric [in {\it index\_pairs}]];\\
\ttoff

Here, {\it index\_pairs} can be {\tt \{i,j\}} etc. but also
\emph{pairs of pairs of indices}, i.e.\ {\tt \{\{i,j\},\{k,l\}\}}
which means that the pair {\tt \{i,j\}} (anti-)commutes with the pair
{\tt \{k,l\}}. The following example will clarify this:

\verbon\begin{verbatim}
index_symmetries t(i,j): symmetric$
index_symmetries r(k,l,m,n): symmetric in {k,l},{m,n} antisymmetric in {{k,l},{m,n}}$
\end{verbatim}
\verboff

These commands mean $t^{ij}=t^{ji}$ and
$r^{klmn}=r^{lkmn}=r^{klnm}=-r^{mnkl}$.

The rest of the commands works rather canonical and can be understood
by looking at some examples: The first example implements the
Schwarzschild metric in the tensor calculus:

\verbon\begin{verbatim}
% file "exc2" 
% The Schwarzschild metric in tensor formalism

load_package excalc$

% introducing the metric (the symbolic coframe is of no importance):
depend(f,r)$ depend(g,r)$
coframe o(0),o(1),o(2),o(3) with metric
g = f**2*o(0)**2 -g**2/f**2*o(1)**2 -r**2*o(2)**2 -r**2*sin(theta)**2*o(3)**2$

% introducing coordinates with running indices:
pform x(i)=0$
x(-0):=t$ x(-1):=r$ x(-2):=theta$ x(-3):=phi$

% calculating the Christoffel symbol:
pform chris(i,j,k)=0$
index_symmetries chris(i,j,k): symmetric in {i,j}$
chris(-i,-j,-k) :=
  (1/2)* (@(g(-j,-k),x(-i)) + @(g(-k,-i),x(-j)) - @(g(-i,-j),x(-k)))$

% the Riemannian curvature:
pform riem(i,j,k,l)=0$
index_symmetries riem(i,j,k,l): antisymmetric in {i,j},{k,l}
                                symmetric in {{i,j},{k,l}}$
riem(-i,-j,-k,l):= 
     @(chris(-j,-k,l),x(-i)) + chris(-i,-m,l)*chris(-j,-k,m)
   - @(chris(-i,-k,l),x(-j)) - chris(-j,-m,l)*chris(-i,-k,m)$

% the Ricci tensor:
pform ric(i,j)=0$
index_symmetries ric(i,j): symmetric$
ric(-i,-j) := riem(-m,-i,-j,m)$

% the Ricci scalar:
pform ricscalar=0$
ricscalar := ric(-i,i);

let {g=>1,f=>sqrt(1-mm/r)}$
ric(-i,-j);

end$
\end{verbatim}
\verboff

The next example proves the Bianchi identity in a purely symbolic manner:

\verbon\begin{verbatim}
% file "exc3"
% Proving the Bianchi identity in the exterior calculus
%
load_package excalc$

indexrange 0,1,2,3$

% general introduction of a connection:
pform gamma(a,b)=1$

% definition of the curvature:
pform curv(a,b)=2$
curv(-a,b):=d gamma(-a,b) - gamma(-a,c) ^ gamma(-c,b)$

% Bianchi identity:
d curv(-a,b) - gamma(-a,c) ^ curv(-c,b) + gamma(-c,b) ^ curv(-a,c);  
  %-> yields zero!

end$
\end{verbatim}
\verboff

The next example implements the Taub-NUT solution with electric charge
in the exterior calculus (in the Kaluza-Klein formalism):

\verbon\begin{verbatim}
% nut.exi, Marc Toussaint, Cologne
% 99-Oct-19
%
% The Taub-NUT solution with electric charge

load_package excalc$ 
off gcd,exp$

% COFRAME %%%%%%%%%%%%%%%%%%%%%%%%%%%%%%%%%%%%%%%%%%%%%%%%%%%%%%%%%%%%%%%%%%%%%

clear t,r,theta,phi,o,e,f,rho,dtau,dsigma,m,n,q$

pform f=0,rho=0$
fdomain f=f(r),rho=rho(r)$
let { dtau   => d t - 2*n*cos(theta)*d phi,
      dsigma => (r**2+n**2)*d phi }$

dim:=5$
coframe o(0) = f/rho* dtau,
        o(1) = rho/f* d r,
        o(2) =  rho * d theta,
        o(3) = 1/rho*sin(theta)* dsigma,
        o(5) = d w + r/rho**2* q*dtau
with signature (1,-1,-1,-1,-1)$
frame e$


% ANSATZ %%%%%%%%%%%%%%%%%%%%%%%%%%%%%%%%%%%%%%%%%%%%%%%%%%%%%%%%%%%%%%%%%%%%%%

f    := sqrt(r**2-2*r*m-n**2+q**2/4)$
rho  := sqrt(r**2+n**2)$


% FIELD EQUATION %%%%%%%%%%%%%%%%%%%%%%%%%%%%%%%%%%%%%%%%%%%%%%%%%%%%%%%%%%%%%%

% Christoffel symbol:
        pform chris1(a,b)=1$
        Riemannconx chris1$
        chris1(a,b):=chris1(b,a)$

% Riemannian curvature and Einstein 3-form:
        pform rie2(a,b)=2,einstein3(a)=dim-1$
        rie2(-a,b):=d chris1(-a,b) + chris1(-c,b)^chris1(-a,c)$
        einstein3(a):=1/2*#(o(a)^o(b)^o(c)) ^ rie2(-b,-c)$

% Result:
        einstein3(a);   %-> zero (except the (5,5)-compopnent)

end;
\end{verbatim}
\verboff

\section{The tensor calculus in Reduce}

The tensor calculus is realized in Reduce by manipulating
\emph{arrays} with elementary commands. Usually, the first step is to
introduce countable (i.e.\ indexed) coordinates as an operator (giving
the coordinate for each index). Next one can introduce a metric and
calculate the inverse to be able to raise and lower indices. Then
tensors are introduced as arrays and all operations are performed by
explicitly running over the indices (with the {\tt for} command).
Whether the array contains covariant or contravariant components is
specified by adding the characters '{\tt l}' for lower and '{\tt u}'
for upper indices to the name. The following example will follow this
scheme to calculate the curvature of a general static, symmetric
geometry in polar coordinates and thereby derive the Schwarzschild
metric.

\verbon\begin{verbatim}
% indexed polar coordinates:
operator x$
x(0):=t$ x(1):=r$ x(2):=theta$ x(3):=phi$

% the metric (gll) with two lower and its inverse (guu) with two upper indices:
array gll(3,3),guu(3,3)$            % arrays are initialized with zero

% static, spherically symmetric ansatz, signature: + - - - :
depend(f,r)$
depend(g,r)$
gll(0,0):=f**2$ gll(1,1):=- g**2/f**2$ gll(2,2):=-r**2$ gll(3,3):=-r**2*(sin(theta))**2$

% calculating the inverse by using Reduce's matrix calculus:
matrix m(4,4)$
for i:=0:3 do for j:=i:3 do m(j+1,i+1):=m(i+1,j+1):=gll(i,j)$
m:=1/m$                         % calculates the inverse
for i:=0:3 do for j:=i:3 do guu(j,i):=guu(i,j):=m(i+1,j+1)$
clear m$

% calculating the Christoffel symbol of type low-low-low and low-low-up:
array chrislll(3,3,3),chrisllu(3,3,3)$
for i:=0:3 do for j:=i:3 do <<
  for k:=0:3 do chrislll(j,i,k) := chrislll(i,j,k) :=
    (1/2) * (df(gll(j,k),x(i)) + df(gll(k,i),x(j)) - df(gll(i,j),x(k)))$
  for k:=0:3 do chrisllu(j,i,k) := chrisllu(i,j,k) :=
    for m:=0:3 sum guu(k,m) * chrislll(i,j,m)$
>>$

% calculating the Riemannian curvature (neglecting symmetries -> inefficient!):
array riemlllu(3,3,3,3)$
for i:=0:3 do for j:=0:3 do for k:=0:3 do for l:=0:3 do
  riemlllu(i,j,k,l):=
    df(chrisllu(j,k,l),x(i)) - df(chrisllu(i,k,l),x(j))
    + for m:=0:3 sum (  chrisllu(i,m,l)*chrisllu(j,k,m)
                      - chrisllu(j,m,l)*chrisllu(i,k,m) )$

% the Ricci tensor:
array ricll(3,3)$
for i:=0:3 do for j:=0:3 do 
  ricll(i,j) := for k:=0:3 sum riemlllu(k,i,j,k)$

% the Ricci scalar:
ricscalar := for k:=0:3 sum for l:=0:3 sum guu(l,k)*ricll(k,l)$

{ricll(0,0),ricll(1,1),ricll(2,2),ricll(3,3)} where {g=>1,f=>sqrt(1-mm/r)};
\end{verbatim}
\verboff


\begin{thebibliography}{99}\thispagestyle{empty}
\bibitem{hehl} F.W.\ Hehl, V.\ Winkelmann, H.\ Meyer: \emph{Reduce: Ein Kompaktkurs \"uber die Anwendung von Computer-Algebra.} Springer, Berlin, 2. Auflage (1992).

\bibitem{rayna} G.\ Rayna: \emph{Reduce: Software for algebraic computation.} Springer, Berlin (1987).

\bibitem{maccallum} M.\ MacCallum, F.\ Wright: \emph{Algebraic computing with Reduce.} Lecture notes from the 1st Brazil school on computer algebra. Clarendon Press, Oxford (1991).

\bibitem{redfern} D.\ Redfern: \emph{The Maple handbook. Maple V release 3.} Springer, Berlin (1994).

\bibitem{mypage} Sources for these lectures: \verb+http://www.thp.uni-koeln.de/~mt/work/1999mexico/+

\bibitem{reducepage} Reduce user's manual: \verb+http://www.uni-koeln.de/REDUCE/3.6/doc/reduce/+

\bibitem{excalcpage} Excalc user's manual: \verb+http://www.uni-koeln.de/REDUCE/3.6/doc/excalc/excalc.html+

\bibitem{maplepage} Maple's homepage: \verb+http://www.maplesoft.com/+

\bibitem{zibpackages} ZIB, Berlin: \verb+http://www.zib.de/Symbolik/reduce/moredocs/+

\bibitem{cathodepage} The CATHODE project: homepage: \verb+http://www-lmc.imag.fr/cathode2/+, software page: \verb+http://www-sop.inria.fr/cafe/Manuel.Bronstein/cathode/software.html+

\end{thebibliography}
\end{document}